\documentclass[journal]{IEEEtran}
\usepackage[dvips]{graphicx,color}
\usepackage{latexsym}
\usepackage{amssymb}
\usepackage{amsmath,bm}
\usepackage{array}

\begin{document}

%
%
\newcommand{\qed}{\hfill$\square$}
\newcommand{\suchthat}{\mbox{~s.t.~}}
\newcommand{\markov}{\leftrightarrow}
\newenvironment{pRoof}{%
 \noindent{\em Proof.\ }}{%
 \hspace*{\fill}\qed \\
 \vspace{2ex}}


\newcommand{\ket}[1]{| #1 \rangle}
\newcommand{\bra}[1]{\langle #1 |}
\newcommand{\bol}[1]{\mathbf{#1}}
\newcommand{\rom}[1]{\mathrm{#1}}
\newcommand{\san}[1]{\mathsf{#1}}
\newcommand{\mymid}{:~}
\newcommand{\argmax}{\mathop{\rm argmax}\limits}
\newcommand{\argmin}{\mathop{\rm argmin}\limits}

\newcommand{\Cls}{class NL}
\newcommand{\vSpa}{\vspace{0.3mm}}
\newcommand{\Prmt}{\zeta}
\newcommand{\pj}{\omega_n}

\newfont{\bg}{cmr10 scaled \magstep4}
\newcommand{\bigzerol}{\smash{\hbox{\bg 0}}}
\newcommand{\bigzerou}{\smash{\lower1.7ex\hbox{\bg 0}}}
\newcommand{\nbn}{\frac{1}{n}}
\newcommand{\ra}{\rightarrow}
\newcommand{\la}{\leftarrow}
\newcommand{\ldo}{\ldots}
\newcommand{\typi}{A_{\epsilon }^{n}}
\newcommand{\bx}{\hspace*{\fill}$\Box$}
\newcommand{\pa}{\vert}
\newcommand{\ignore}[1]{}

%
%
%
%
\newcommand{\bc}{\begin{center}}  %
\newcommand{\ec}{\end{center}}
\newcommand{\befi}{\begin{figure}[h]}  %
\newcommand{\enfi}{\end{figure}}
\newcommand{\bsb}{\begin{shadebox}\begin{center}}   %
\newcommand{\esb}{\end{center}\end{shadebox}}
\newcommand{\bs}{\begin{screen}}     %
\newcommand{\es}{\end{screen}}
\newcommand{\bib}{\begin{itembox}}   %
\newcommand{\eib}{\end{itembox}}
\newcommand{\bit}{\begin{itemize}}   %
\newcommand{\eit}{\end{itemize}}
\newcommand{\defeq}{:=}
\newcommand{\Qed}{\hbox{\rule[-2pt]{3pt}{6pt}}}
\newcommand{\beq}{\begin{equation}}
\newcommand{\eeq}{\end{equation}}
\newcommand{\beqa}{\begin{eqnarray}}
\newcommand{\eeqa}{\end{eqnarray}}
\newcommand{\beqno}{\begin{eqnarray*}}
\newcommand{\eeqno}{\end{eqnarray*}}
\newcommand{\ba}{\begin{array}}
\newcommand{\ea}{\end{array}}
\newcommand{\vc}[1]{\mbox{\boldmath $#1$}}
\newcommand{\lvc}[1]{\mbox{\scriptsize \boldmath $#1$}}
\newcommand{\svc}[1]{\mbox{\scriptsize\boldmath $#1$}}

\newcommand{\wh}{\widehat}
\newcommand{\wt}{\widetilde}
\newcommand{\ts}{\textstyle}
\newcommand{\ds}{\displaystyle}
\newcommand{\scs}{\scriptstyle}
\newcommand{\vep}{\varepsilon}
\newcommand{\rhp}{\rightharpoonup}
\newcommand{\cl}{\circ\!\!\!\!\!-}
\newcommand{\bcs}{\dot{\,}.\dot{\,}}
\newcommand{\eqv}{\Leftrightarrow}
\newcommand{\leqv}{\Longleftrightarrow}
\newtheorem{co}{Corollary} 
\newtheorem{lm}{Lemma} 
\newtheorem{Ex}{Example} 
\newtheorem{Th}{Theorem}
\newtheorem{df}{Definition} 
\newtheorem{pr}{Property} 
\newtheorem{pro}{Proposition} 
\newtheorem{rem}{Remark} 
\newtheorem{assm}{Assumption} 

\newcommand{\lcv}{convex } 
\newcommand{\hugel}{{\arraycolsep 0mm
                    \left\{\ba{l}{\,}\\{\,}\ea\right.\!\!}}
\newcommand{\Hugel}{{\arraycolsep 0mm
                    \left\{\ba{l}{\,}\\{\,}\\{\,}\ea\right.\!\!}}
\newcommand{\HUgel}{{\arraycolsep 0mm
                    \left\{\ba{l}{\,}\\{\,}\\{\,}\vspace{-1mm}
                    \\{\,}\ea\right.\!\!}}
\newcommand{\huger}{{\arraycolsep 0mm
                    \left.\ba{l}{\,}\\{\,}\ea\!\!\right\}}}
\newcommand{\Huger}{{\arraycolsep 0mm
                    \left.\ba{l}{\,}\\{\,}\\{\,}\ea\!\!\right\}}}
\newcommand{\HUger}{{\arraycolsep 0mm
                    \left.\ba{l}{\,}\\{\,}\\{\,}\vspace{-1mm}
                    \\{\,}\ea\!\!\right\}}}
\newcommand{\hugebl}{{\arraycolsep 0mm
                    \left[\ba{l}{\,}\\{\,}\ea\right.\!\!}}
\newcommand{\Hugebl}{{\arraycolsep 0mm
                    \left[\ba{l}{\,}\\{\,}\\{\,}\ea\right.\!\!}}
\newcommand{\HUgebl}{{\arraycolsep 0mm
                    \left[\ba{l}{\,}\\{\,}\\{\,}\vspace{-1mm}
                    \\{\,}\ea\right.\!\!}}
\newcommand{\hugebr}{{\arraycolsep 0mm
                    \left.\ba{l}{\,}\\{\,}\ea\!\!\right]}}
\newcommand{\Hugebr}{{\arraycolsep 0mm
                    \left.\ba{l}{\,}\\{\,}\\{\,}\ea\!\!\right]}}
\newcommand{\HugebrB}{{\arraycolsep 0mm
                    \left.\ba{l}{\,}\\{\,}\vspace*{-1mm}\\{\,}\ea\!\!\right]}}
\newcommand{\HUgebr}{{\arraycolsep 0mm
                    \left.\ba{l}{\,}\\{\,}\\{\,}\vspace{-1mm}
                    \\{\,}\ea\!\!\right]}}
\newcommand{\hugecl}{{\arraycolsep 0mm
                    \left(\ba{l}{\,}\\{\,}\ea\right.\!\!}}
\newcommand{\Hugecl}{{\arraycolsep 0mm
                    \left(\ba{l}{\,}\\{\,}\\{\,}\ea\right.\!\!}}
\newcommand{\HUgecl}{{\arraycolsep 0mm
                    \left(\ba{l}{\,}\\{\,}\\{\,}\vspace{-1mm}
                    \\{\,}\ea\right.\!\!}}
\newcommand{\hugecr}{{\arraycolsep 0mm
                    \left.\ba{l}{\,}\\{\,}\ea\!\!\right)}}
\newcommand{\Hugecr}{{\arraycolsep 0mm
                    \left.\ba{l}{\,}\\{\,}\\{\,}\ea\!\!\right)}}
\newcommand{\HUgecr}{{\arraycolsep 0mm
                    \left.\ba{l}{\,}\\{\,}\\{\,}\vspace{-1mm}
                    \\{\,}\ea\!\!\right)}}
\newcommand{\hugepl}{{\arraycolsep 0mm
                    \left|\ba{l}{\,}\\{\,}\ea\right.\!\!}}
\newcommand{\Hugepl}{{\arraycolsep 0mm
                    \left|\ba{l}{\,}\\{\,}\\{\,}\ea\right.\!\!}}
\newcommand{\hugepr}{{\arraycolsep 0mm
                    \left.\ba{l}{\,}\\{\,}\ea\!\!\right|}}
\newcommand{\Hugepr}{{\arraycolsep 0mm
                    \left.\ba{l}{\,}\\{\,}\\{\,}\ea\!\!\right|}}
\newcommand{\MEq}[1]{\stackrel{
{\rm (#1)}}{=}}

\newcommand{\MLeq}[1]{\stackrel{
{\rm (#1)}}{\leq}}

\newcommand{\ML}[1]{\stackrel{
{\rm (#1)}}{<}}

\newcommand{\MGeq}[1]{\stackrel{
{\rm (#1)}}{\geq}}

\newcommand{\MG}[1]{\stackrel{
{\rm (#1)}}{>}}

\newcommand{\MPreq}[1]{\stackrel{
{\rm (#1)}}{\preceq}}

\newcommand{\MSueq}[1]{\stackrel{
{\rm (#1)}}{\succeq}}

\newenvironment{jenumerate}
	{\begin{enumerate}\itemsep=-0.25em \parindent=1zw}{\end{enumerate}}
\newenvironment{jdescription}
	{\begin{description}\itemsep=-0.25em \parindent=1zw}{\end{description}}
\newenvironment{jitemize}
	{\begin{itemize}\itemsep=-0.25em \parindent=1zw}{\end{itemize}}
\renewcommand{\labelitemii}{$\cdot$}

\newcommand{\iro}[2]{{\color[named]{#1}#2\normalcolor}}

\newcommand{\irr}[1]{{\color[named]{Black}#1\normalcolor}}

\newcommand{\irg}[1]{{\color[named]{Green}#1\normalcolor}}
\newcommand{\irb}[1]{{\color[named]{Blue}#1\normalcolor}}
\newcommand{\irBl}[1]{{\color[named]{Black}#1\normalcolor}}
\newcommand{\irWh}[1]{{\color[named]{White}#1\normalcolor}}

\newcommand{\irY}[1]{{\color[named]{Yellow}#1\normalcolor}}
\newcommand{\irO}[1]{{\color[named]{Orange}#1\normalcolor}}
\newcommand{\irBr}[1]{{\color[named]{Purple}#1\normalcolor}}
\newcommand{\IrBr}[1]{{\color[named]{Purple}#1\normalcolor}}
\newcommand{\irBw}[1]{{\color[named]{Brown}#1\normalcolor}}
\newcommand{\irPk}[1]{{\color[named]{Magenta}#1\normalcolor}}
\newcommand{\irCb}[1]{{\color[named]{CadetBlue}#1\normalcolor}}

\newcommand{\Vcx}{\mbox{\boldmath $x$}}
\newcommand{\Vcy}{\mbox{\boldmath $y$}}

\newcommand{\calVarX}{{\cal X}}
\newcommand{\calVarY}{{\cal Y}}
\newcommand{\calVarZ}{\cal Z}

\newcommand{\calVarXun}{{\cal X}^n}
\newcommand{\calVarYun}{{\cal Y}^n}
\newcommand{\calVarZum}{{\cal Z}^m}

\newcommand{\varx}{x}
\newcommand{\varxun}{x^n}
\newcommand{\varxa}{x_{1}}
\newcommand{\varxb}{x_{2}}
\newcommand{\varxn}{x_{n}}
\newcommand{\varxt}{x_{t}}

\newcommand{\VarXa}{X_{1}}
\newcommand{\VarXb}{X_{2}}
\newcommand{\VarXn}{X_{n}}
\newcommand{\VarXt}{X_{t}}

\newcommand{\tX}{\tilde{X}}
\newcommand{\tY}{\tilde{Y}}
\newcommand{\tZ}{\tilde{Z}}
\newcommand{\tZum}{\tilde{Z}^m}

\newcommand{\bX}{\bar{X}}
\newcommand{\bY}{\bar{Y}}
\newcommand{\bZ}{\bar{Z}}
\newcommand{\bZum}{\bar{Z}^m}

%
\newenvironment{indention}[1]{\par
\addtolength{\leftskip}{#1}\begingroup}{\endgroup\par}
%
\newcommand{\namelistlabel}[1]{\mbox{#1}\hfill} 
\newenvironment{namelist}[1]{%
\begin{list}{}
{\let\makelabel\namelistlabel
\settowidth{\labelwidth}{#1}
\setlength{\leftmargin}{1.1\labelwidth}}
}{%
\end{list}}
%
%
\newcommand{\bfig}{\begin{figure}[t]}
\newcommand{\efig}{\end{figure}}
\setcounter{page}{1}

\newtheorem{theorem}{Theorem}

\newcommand{\ep}{\mbox{\rm e}}

\newcommand{\ExP}{\rm e}
\newcommand{\Ep}{\mbox{\rm e}}

\newcommand{\Exp}{\exp
}
\newcommand{\idenc}{{\varphi}_n}
\newcommand{\resenc}{
{\varphi}_n}
\newcommand{\ID}{\mbox{\scriptsize ID}}
\newcommand{\TR}{\mbox{\scriptsize TR}}
\newcommand{\Av}{\mbox{\sf E}}

\newcommand{\Vl}{|}
\newcommand{\Ag}{(R,P_{X^n}|W^n)}
\newcommand{\Agv}[1]{({#1},P_{X^n}|W^n)}
\newcommand{\Avw}[1]{({#1}|W^n)}

\newcommand{\Dist}{\Delta}

\newcommand{\Jd}{X^nY^n}
\newcommand{\IdR}{r_n}

\newcommand{\Index}{{n,i}}

\newcommand{\cid}{C_{\mbox{\scriptsize ID}}}
\newcommand{\cida}{C_{\mbox{{\scriptsize ID,a}}}}

\newcommand{\OMega}
{\Omega^{(\mu,\lambda,\alpha)}}
\newcommand{\ARgRv}{(p^{(n)},q_{X^n})}

\newcommand{\pmt}{\beta}
\newcommand{\OHCS}{Z}

\arraycolsep 0.5mm
\date{}
%
\title{
A New Inequality Related to Proofs of 
Strong Converse Theorems 
for Source or Channel Networks
}
\author{%
Yasutada Oohama 
\thanks{
Y. Oohama is with 
University of Electro-Communications,
1-5-1 Chofugaoka Chofu-shi, Tokyo 182-8585, Japan.
}%
}
\markboth{
}
{
}
\maketitle

\begin{abstract} 
In this paper we provide a new inequality useful for 
the proofs of strong converse theorems in the 
multiterminal information theory. We apply this 
inequality to the recent work by Tyagi and Watanabe 
on the strong converse theorem for the Wyner-Ziv 
source coding problem to obtain a new strong 
converse outer bound. This outer bound deviates 
from the Wyner-Ziv rate distortion 
region with the order $O\left(\frac{1}{\sqrt{n}}\right)$ 
on the length $n$ of source outputs. 
\end{abstract}

\section{Definitions of Functions} 

Let $\Lambda\defeq \{1,2,\cdots,m \}$ be an index set. 
For each $i\in {\Lambda }$, let ${{\cal X}}_i$ be a finite set. 
For each $i\in {\Lambda }$, let ${X}_i$ be a random variable 
taking values in ${{\cal X}}_i$. 
For $S\subseteq \Lambda$, ${X}_S\defeq ({X}_i)_{i\in S}$. 
In particular for $S=\Lambda$, we write 
${X}_{\Lambda}=\underline{X}$. 
Let ${\cal P}$ be a set of all probability distributions 
on $\underline{{\cal X}}$. 
For $\underline{X} \in \underline{{\cal X}}$, 
we write its disribution as 
$p=p_{\underline{X}}\in {\cal P}$. For $p_{\underline{X}}$,
we often omit its subscript $\underline{X}$ to simply write 
$p$. For $S \subseteq \Lambda$, let 
$
p_{{X}_S}=\{p_{{X}_S}(x_S)\}_{x_S \in {{\cal X}}_S}
$
denote the probability distribution of ${X}_S$, 
which is the marginral distribution of $p \in {\cal P}$. 
We adopt similar notations for other variables or sets. 
For $p \in {\cal P}$, we consider a function 
$\omega_p(\underline{x}),\underline{x}\in {{\cal X}}$ 
having the following form:
\begin{align}
&\omega_p(\underline{x})
= \sum_{l=1}^{L_0} \xi_l\phi_l(\underline{x})
\label{eqn:DefOne}
\\
& \quad
  +\sum_{l=1}^{L_1}  \mu_l \log p_{{X}_{S_l}}(x_{S_l}) 
  -\sum_{l=1}^{L_2} \eta_l \log p_{{X}_{T_l}}(x_{T_l}). 
\label{eqn:DefTwo}
\end{align}
In (\ref{eqn:DefOne}), $\phi_l(\underline{x}), 
\underline{x}\in \underline{\cal X}$, 
$l=1,2,\cdots, L_0$, 
are given $L_0$ nonnegative functions 
and $\xi_l, l=1,2, \cdots,L_0$, 
are given $L_0$ real valued coefficients.  
In (\ref{eqn:DefTwo}), the quantities 
$\mu_l, l=1,2, \cdots,L_1$ and  
$\eta_l,l=1,2, \cdots,L_2$ 
are given $(L_1+L_2)$ positive coefficients. 
Furthermore, 
$S_l,l=1,2,\cdots, L_1$ and 
$T_l,l=1,2,\cdots, L_2$ 
are given $(L_1+L_2)$ subsets of $\Lambda$. 
We define
\begin{align}
\tilde{\Psi} (p) \defeq {\rm E}_p \left[ \omega_p(\underline{X}) \right]
=\sum_{\underline{x}\in \underline{{\cal X}}}
p(\underline{x})\omega_p(\underline{x}).
\end{align}
In this paper we assume that the function 
$\omega_p= \left\{ \omega_p(\underline{x})
\right\}_{\underline{x}\in \underline{{\cal X}}}$ 
satisfy the following property.
\begin{assm}\label{assm:assmA} $\quad$
\begin{itemize}

\item[a)] For any $p \in {\cal P}$, $\tilde{\Psi}(p)$ 
is nonnegative and bounded, i.e., there exists 
a positive $K$ such that $\tilde{\Psi}(p)$ $\in [0,K]$ 
for any $p\in {\cal P}$.

\item[b)] $\tilde{\Psi}(p)$ is a continuous function of $p \in {\cal P}$. 
\end{itemize}
\end{assm}

Let $\tilde{\cal P}$ be a given subset of ${\cal P}$. 
The following two optimization problems 
\begin{align}
\tilde{\Psi}_{\max}& \defeq \max_{p \in \tilde{P}}\Psi(p) 
\mbox{ or }
\tilde{\Psi}_{\min} = \min_{p \in \tilde{P}}\Psi(p) 
\label{eqn:OptB}
\end{align}
frequently appear in the analysis of capacity or rate regions 
in the field of multiterminal information theory. In this paper 
we give one example of $\omega_p(\underline{X})$ and $\tilde{\cal P}$, 
which is related to the source coding with side 
information at the deconder posed and investigated Wyner 
and Ziv \cite{wz}. 
This example is shown below.

\begin{Ex}\label{ex:ExOne}
Let $U$, $X$, $ Y$, and $Z$ 
be four random variables, respectively taking values 
in the finite sets ${\cal U}$, ${\cal X}$, ${\cal Y}$, 
and ${\cal Z}$. 
We consider the case where $\underline{X}=(U,X,Y,Z)$.
Let $p=p_{UXYZ}$ a  probability distribution of $(U,X,Y,Z)$.
For $(u,x,y,z)\in $
${\cal U}\times $ 
${\cal X}\times $ 
${\cal Y}\times $ 
${\cal Z}$, 
we define
\begin{align}
& \omega_p(u,x,y,z) \defeq 
  \xi d(x,z) + \bar{\xi}\log \frac{p_{X|UY}(x|u,y)}{p_{X|Y}(x|y)}
\notag\\
&=\xi d(x,z) + \bar{\xi}[\log p_{UXY}(u,x,y)+ \log p_{Y}(y)]
\notag\\
&\quad        -\bar{\xi}[\log p_{UY}(u,y)+ \log p_{XY}(x,y)],
\end{align}
where $d(x,z),(x,z) \in {\cal X} \times {\cal Z}$ are 
distortion measures. In this example we have 
the following:
\beq
\left.
\ba{ll}
L_1=2,\mu_1=\mu_2=\bar{\xi},   &S_1=\{U,X,Y\},S_2=\{Y\}, 
\\
L_2=2,\eta_1=\eta_2=\bar{\xi}, &T_1=\{U,Y\},T_2=\{X,Y\}.
\ea
\right\}
\eeq
Let 
\begin{align*}
& \tilde{\cal P}=\tilde{\cal P}(p_{XY})
=\{q=q_{UXYZ}:|{\cal U}|\leq |{\cal X}|,q_{XY}=p_{XY} 
\\
&\quad U \markov  X \markov Y,\: X \markov (U,Y) \markov Z \}. 
\end{align*} 
In this example we  denote the quantity 
$\tilde{\Psi}_{\min}$ by $R_{\sf WZ}^{(\xi)}(p_{XY})$, which 
has the following form:
\begin{align*}
R_{\sf WZ}^{(\xi)}(p_{XY})=
\min_{q \in \tilde{\cal P}}\left[ \bar{\xi} I_q(X;U|Y) + \xi {\rm E}_q d(X,Z) \right].
\end{align*}
The quantity $R_{\sf WZ}^{(\xi)}(p_{XY})$ yields 
the following hyperplane expression of Wyner-Ziv rate 
distortion region ${\cal R}_{\sf WZ}(p_{XY})$: 
$$
{\cal R}_{\sf WZ}(p_{XY})=
\bigcap_{\xi \in[0,1]}\{(R,D): 
\bar{\xi} R+\xi D \geq R_{\sf WZ}^{(\xi)}(p_{XY})\}. 
$$
\end{Ex}

In the above example because of the two Markov chains  
$U \markov X \markov Y$ and $X \markov (U,Y) \markov Z$, 
the computation of $R_{\sf WZ}^{(\xi)}(p_{XY})$ becomes 
a non-convex optimization problem, which is very hard 
to solve in its present form. As we can see from this 
example, the computations of $\tilde{\Psi}_{\min}$ and 
$\tilde{\Psi}_{\max}$ are in general highly challenging. 
To solve those problems, alternative optimization problems 
having one parameter on some relaxed condition of $\tilde{P}$ 
are introduced. Let 
$\varphi: {\cal P} \to \tilde{\cal P}$ be some suitable onto 
mapping satisfying 
$\varphi(q)=q \mbox{ if }q \in \tilde{\cal P}.$ 
We set 
$
p=p^{(q)}\defeq \varphi(q).   
$
On the above $\varphi$, we assume the following:
\begin{assm}
\label{asm:AssmTwo}$\quad$
\begin{itemize}
\item[a)]  
Let ${\cal P}^{*}$ denote 
a feasible region ${\cal P}^{*}$ on those relaxed optimization problems.
On the feasible region ${\cal P}^{*}$, we assume that for any 
$q \in {\cal P}^{*}$, its support set ${\rm Supp}(q)$ includes the 
support set ${\rm Supp}(\varphi(q))$ of $\varphi(q)$. 

\item[b)]
For any $q \in {\cal P}^{*}$ and for any 
$p=p^{(q)} \in \tilde{\cal P}$, we have   
\begin{align*}
\!\!\!\!& \omega_q(\underline{x})-\omega_p(\underline{x})
=\sum_{l=1}^{L_3} \kappa_l \log 
\frac{ p_{X_{{A}_{2l}} | X_{{A}_{2l-1}}}
         (x_{{A}_{2l}} | x_{{A}_{2l-1}}) }
     { q_{X_{{A}_{2l}} | X_{{A}_{2l-1}}}
         (x_{{A}_{2l}} | x_{{A}_{2l-1}}) }
\notag\\
\!\!\!\!&\quad -\sum_{l=1}^{L_4} \nu_l \log 
\frac{ p_{X_{B_{2l}} | X_{B_{2l-1}}}
         (x_{B_{2l}} | x_{B_{2l-1}}) }
     { q_{X_{B_{2l}} | X_{B_{2l-1}}}
         (x_{B_{2l}} | x_{B_{2l-1}}) },
\end{align*}
where $\kappa_l,l=1,2,\cdots, L_3$ 
and $\nu_l,l=1,2, \cdots,L_4$ are $(L_3+L_4)$ positive 
constants and the quantities 
$\{ {A}_{2l-1},{A}_{2l} \}_{l=1}^{L_3}$ and 
$\{ B_{2l-1},B_{2l} \}_{l=1}^{L_4}$ 
are $2(L_3+L_4)$ subsets of $\Lambda$ 
satistying the following:
\begin{align*}
& {A}_{2l-1}\cap {A}_{2l}=\emptyset \mbox{ for }l=1,2,\cdots, L_3, 
\\
& B_{2l-1}\cap B_{2l}=\emptyset \mbox{ for }l=1,2,\cdots, L_4.  
\end{align*}
\end{itemize}
\end{assm}

For $\alpha>0$ and $q \in {\cal P}^{*}$, define
\begin{align*}
\Psi^{(\alpha)}(q)
&\defeq       {\rm E}_q \left[ \omega_q(\underline{X})
     -\alpha \log \frac{q(\underline{X})}{p^{(q)}(\underline{X})}\right]
\\
&=    {\rm E}_q \left[ \omega_q(\underline{X})\right]-\alpha 
      D\left(q\Bigl|\Bigr|p^{(q)}\right).
\end{align*}
We consider the following two optimization problems:
\begin{align}
  {\Psi}_{\max}^{(\alpha)}& \defeq 
\max_{q \in {\cal P}^{*}}{\Psi}^{(\alpha)}(q) 
\mbox{ or }
\Psi_{\min}^{(-\alpha)} = \min_{q \in {\cal P}^{*}}
{\Psi}^{(-\alpha)}(q). 
\label{eqn:OptD}
\end{align}
Those optimization problems 
appear in recent results that 
the author \cite{OhIsit15AKWStConv}-\cite{OhABCArXiv16}, 
Tyagi and Watanabe \cite{TyagiWatanabeArXiv18} obtained on 
the proofs of the strong converse theorems for multi-terminal 
source or channel networks. 

\begin{Ex}\label{ex:ExTwo} 
We consider the case of Example \ref{ex:ExOne}. 
Define $\varphi:{\cal P}\to \tilde{\cal P}$ by 
$
\varphi(q)=\tilde{q}=(q_{U|X},p_{XY},q_{Z|UY}) \in \tilde{\cal P}.
$
The feasible region ${\cal P}^{*}\subseteq {\cal P}$ is given by 
\begin{align*}
{\cal P}^{*}=&\{q=q_{UXYZ}:|{\cal U}|\leq |{\cal X}||{\cal Y}||{\cal Z}|, 
\\
&\:{\rm Supp}(q) \supseteq {\rm Supp}(\varphi(q))\}.
\end{align*}
For $q \in {\cal P}^{*}$ and for 
$\tilde{q}=\tilde{q}^{(q)}=\varphi(q)$, we have 
\begin{align}
&\omega_q(u,x,y,z)-\omega_{\tilde{q}}(u,x,y,z)
\notag\\
&=      \bar{\xi}\log \frac{\tilde{q}_{U|Y}(u|y)}
                           {q_{U|Y}(u|y)}
       -\bar{\xi}\log \frac{\tilde{q}_{U|XY}(u|x,y)}
                           {q_{U|XY}(u|x,y)}.
\label{eqn:AdssR}
\end{align}
From (\ref{eqn:AdssR}), we have that 
$L_3=1,\kappa_1=\bar{\xi},$ and $L_4=1,\nu_1=\bar{\xi}$.
We denote the quantity ${\Psi}_{\min}^{(-\alpha)}$ by   
$R_{\sf WZ}^{(\xi,\alpha)}(p_{XY})$, which has 
the following form:
\begin{align*}
&R_{\sf WZ}^{(\xi,\alpha)}(p_{XY})
\\
&=\min_{q\in {\cal P}^{*}}\left[\bar{\xi} I_q(X;U|Y) + \xi{\rm E}_q d(X,Z) 
  +\alpha D(q||\varphi(q))\right]
\\
&=\min_{q \in {\cal P}^{*}} \Bigl[\bar{\xi}I_q(X;U|Y)+ \xi {\rm E}_q d(X,Z) 
     +\alpha \{I_q(Y;U|X)
\\
& \qquad +D(q_{XY}||p_{XY})+I_q(X;Z|U,Y)\}\Bigr].
\end{align*}
According to Tyagi and Watanabe \cite{TyagiWatanabeArXiv18}, 
a single letter characterization of the rate
distortion region using the function 
$R_{\sf WZ}^{(\xi,\alpha)}(p_{XY})$ plays an important 
role in the proof of the the strong converse theorem 
for Wyner-Ziv source coding problem.
\end{Ex}

\newcommand{\bibDatA}{
\bibitem{OhIsit15AKWStConv}
Y. Oohama, ``Exponent function for one helper source coding problem 
at rates outside the rate region,'' 
{\it Proceedings of the 2015 IEEE International
Symposium on Information Theory}, pp. 1575--1579, Hong Kong, China, 
June 14-19, 2015, the extended version 
is available at https://arxiv.org/pdf/1504.0589.pdf.



\bibitem{OhEnt18wz}
Y. Oohama, ``Exponential strong converse for 
source coding with side information at the 
decoder,'' {\it Entropy 2018}, 20(5), 
352; doi:10.3390/e20050352.

\bibitem{OhIsita16abc}
Y. Oohama, ``Exponent function for asymmetric broadcast
channels at rates outside the capacity region,''
{\it Proceedings of the 2016 IEEE InternationalSymposium on Information 
Theory and its Applications}, pp. 568--572, 
Monterey, USA, Oct. 30-Nov. 2, 2016.

\bibitem{OhABCArXiv16}
Y. Oohama, ``New strong converse for asymmetric 
broadcast channels,''
{\it preprint;} available at 
https://arxiv.org/pdf/1604.02901.pdf.

\bibitem{TyagiWatanabeArXiv18}
H. Tyagi and S. Watanabe, 
``Strong converse using change of measure arguments,''
{\it preprint;} available at 
https://arxiv.org/pdf/1805.04625.pdf.

\bibitem{wz}A. D. Wyner and J. Ziv, 
``The rate-distortion function for source coding with side 
information at the decoder,''
{\it IEEE Trans. Inform. Theory}, vol. IT-22, pp. 1-10, Jan. 1976.  


\bibitem{han}
T. S. Han, {\it Information-Spectrum Methods in Information
Theory. }Springer-Verlag, Berlin, New York, 2002. The Japanese 
edition was published by Baifukan-publisher, Tokyo, 1998.

\bibitem{ck} 
I. Csisz\'ar and J. K\"orner, 
{\it Information Theory: Coding Theorems for Discrete 
Memoryless Systems.} 
{the second edition, Cambridge University Press, 2011.}


}

\section{
Main Results
}

Our aim in this paper is to evaluate the differences between
$\tilde{\Psi}_{\max}$ and ${\Psi}^{(\alpha)}_{\max}$ 
and between  $\tilde{\Psi}_{\min}$ and ${\Psi}^{(\alpha)}_{\min}$. 
It is obvious that we have 
\beq
\tilde{\Psi}_{\max} \leq \Psi_{\max}^{(\alpha)},\quad
\tilde{\Psi}_{\min} \geq \Psi_{\min}^{(-\alpha)}.
\label{eqn:ZxCc}
\eeq
for any $\alpha \geq 0$. In fact, restricting the feasible region 
${\cal P}^{*}$ in the definitions of $\Psi_{\max}^{(\alpha)}$
or $\Psi_{\min}^{(\alpha)}$ to $\tilde{\cal P}$, we obtain 
the bounds in (\ref{eqn:ZxCc}).
We first describe explicit upper bounds of 
$\Psi_{\max}^{(\alpha)}-\tilde{\Psi}_{\max}$ 
and $\tilde{\Psi}_{\min}-\Psi_{\min}^{(-\alpha)}$
by standard analytical arguments. This result is given by 
the following proposition.

\begin{pro}\label{pro:ProOne} 
For any positive $\alpha$, we have
\begin{align} 
&0\leq \Psi_{\max}^{(\alpha)}-\tilde{\Psi}_{\max}
\leq {K}^{\prime}\sqrt{\frac{2K}{\alpha}} 
\log \left(\sqrt{\frac{\alpha }{2K}}{{\rm e}|\underline{{\cal X}}|}\right),
\label{eqn:ProOneIeq}
\\
&0\leq \tilde{\Psi}_{\min} - \Psi_{\min}^{(-\alpha)}
\leq {K}^{\prime}\sqrt{\frac{2K}{\alpha}} 
\log \left(\sqrt{\frac{\alpha }{2K}}{{\rm e}|\underline{{\cal X}}|}\right),
\label{eqn:ProOneIeqX}
\end{align}
where we set 
\begin{align*}
& \phi_{\max}
\defeq \max_{1\leq l\leq L_1}
       \max_{\underline{x}\in \underline{{\cal X}}}
       \phi_{l}(\underline{x}), 
\\
& {K}^{\prime} \defeq \max\left\{
\phi_{\max}\sum_{l=1}^{L_0}|\xi_l|,
\sum_{l=1}^{L_1} |\mu_l|,
\sum_{l=1}^{L_2}|\eta_l|
\right\}.  
\end{align*}
\end{pro}

Proof of this proposition is given in Appendix \ref{sub:ApdProOne}.
\newcommand{\ApdaProOne}{
\subsection{
Proof of Proposition \ref{pro:ProOne}
}\label{sub:ApdProOne}

In this appendix we prove Proposition \ref{pro:ProOne}.

{\it Proof of Proposition \ref{pro:ProOne}: }
We first examine an upper bound of 
$\Psi_{\max}^{(\alpha)}-\tilde{\Psi}_{\max}$.
Let $q_{\rm opt}^{(\alpha)}\in {\cal P}$ 
be a distribution that attains the maximum of 
${\Psi}^{(\alpha)}_{\max}$. For 
$S \subseteq \Lambda$, 
$q_{{\rm opt},{X}_{S}}^{(\alpha)}$ stands for a marginal 
distribution of $q_{\rm opt}^{(\alpha)}$.   
Then we have the following:  
\begin{align}
&\alpha D(q_{\rm opt}^{(\alpha)}||p^{(q_{\rm opt}^{(\alpha)})})
=\tilde{\Psi}\left(q_{\rm opt}^{(\alpha)}\right)
      -{\Psi}_{\max}^{(\alpha)}  
\MLeq{a} K.
\label{eqn:AssPz} 
\end{align}
Step (a) follows from Assumption \ref{assm:assmA}
part a). From (\ref{eqn:AssPz}), we have
$$
D(q_{\rm opt}^{(\alpha)}||p^{( q_{\rm opt}^{(\alpha)}) })
\leq \frac{K}{\alpha}.
$$
Then, for any $l=1,2,\cdots,L$, we have 
the following chain of inequalities:
\begin{align}
&\sum_{ {a}_{S_l} \in {{\cal X}}_{S_l}}
\left|q_{{\rm opt}, {X}_{S_l}}^{(\alpha)}({a_{S_l}})
 -p_{{X}_{S_l}}^{(q_{\rm opt}^{(\alpha)})}({a_{S_l}})
\right|
\notag\\
&\leq \sum_{\underline{x}\in \underline{{\cal X}}}
\left|
     q_{\rm opt}^{(\alpha)}(\underline{x})
-p^{(q_{\rm opt}^{(\alpha)})}(\underline{x})
\right|
\MLeq{a} \sqrt{2D(q_{\rm opt}^{(\alpha)}||p^{(q_{\rm opt}^{(\alpha)})})}
\notag\\
&\leq \sqrt{\frac{2K}{\alpha}}.
\label{eqn:Ieqa}
\end{align}
Step (a) follows from the Pinsker's inequality.
On upper bound of 
${\Psi}_{\max}^{(\alpha)}-\tilde{\Psi}_{\max}$, 
we have the following chain of inequalities: 

\begin{align}
& \Psi_{\max}^{(\alpha)}-\tilde{\Psi}_{\max}
\notag\\
& = \tilde{\Psi}\left(q_{\rm opt}^{(\alpha)}\right)
  - \alpha D(q_{\rm opt}^{(\alpha)}||p^{q_{\rm opt}^{(\alpha)}})
  -\tilde{\Psi}_{\max}
\notag\\
& \leq \tilde{\Psi}\left(q_{\rm opt}^{(\alpha)}\right)-\tilde{\Psi}_{\max}
\notag\\
&=\tilde{\Psi}\left(p^{(q_{\rm opt}^{(\alpha)})}\right)-\tilde{\Psi}_{\max}
     +\tilde{\Psi}\left(q_{\rm opt}^{(\alpha)}\right)
  -\tilde{\Psi}\left(p^{(q_{\rm opt}^{(\alpha)})}\right) 
\notag\\
&\leq \tilde{\Psi}\left(q_{\rm opt}^{(\alpha)}\right)
  -\tilde{\Psi}\left(p^{(q_{\rm opt}^{(\alpha)})}\right). 
\label{eqn:SdZx}
\end{align}
From (\ref{eqn:SdZx}), we have the following chain of inequalities:  
\begin{align}
& \Psi_{\max}^{(\alpha)}-\tilde{\Psi}_{\max}
\notag\\
&\leq  \sum_{\underline{x}\in \underline{{\cal X}}}
   \left[q_{\rm opt}^{(\alpha)}(\underline{x})
 \omega_{q_{\rm opt}^{(\alpha)}}(\underline{x})
        -p^{(q_{\rm opt}^{(\alpha)})}(\underline{x})
 \omega_{p^{(q_{\rm opt}^{(\alpha)})}}(\underline{x})
\right]
\notag\\
&\leq  \sum_{l=1}^{L_0}|\xi_l|\sum_{\underline{x}\in \underline{{\cal X}}}
\phi_l(\underline{x})
\left|q_{\rm opt}^{(\alpha)}(\underline{x})-  
  p^{(q_{\rm opt}^{(\alpha)})}(\underline{x})\right|
\notag\\
&\quad + \sum_{l=1}^{L_1} |\mu_l| 
     \sum_{a_{S_l} \in {{\cal X}}_{S_l}} 
     \biggl| q_{\rm opt}^{(\alpha)}(\underline{x}_{S_l})
        \log q_{\rm opt}^{(\alpha)}(\underline{x}_{S_l})
\notag\\
&\qquad \qquad \qquad \quad
              -p^{(q_{\rm opt}^{(\alpha)})}({a}_{S_l})
          \log p^{(q_{\rm opt}^{(\alpha)})}({a}_{S_l})\biggr|
\notag\\
&\quad + \sum_{l=1}^{L_2} |\eta_l| 
     \sum_{a_{T_l} \in {{\cal X}}_{T_l}} 
     \biggl| q_{\rm opt}^{(\alpha)}(\underline{x}_{T_l})
        \log q_{\rm opt}^{(\alpha)}(\underline{x}_{T_l})
\notag\\
&\qquad \qquad \qquad \quad
              -p^{(q_{\rm opt}^{(\alpha)})}({a}_{T_l})
          \log p^{(q_{\rm opt}^{(\alpha)})}({a}_{T_l})\biggr|
\notag\\
&\MLeq{a} 
\sqrt{\frac{2K}{\alpha}}
\left[
   \phi_{\max} \sum_{l=1}^{L_0} |\xi_l| 
 + \sum_{l=1}^{L_1}|\mu_l| 
  \log \left(\sqrt{\frac{\alpha }{2K}} |{{\cal X}}_{S_l}|\right)
\right.
\notag\\
& \left.
   \quad + \sum_{l=1}^{L_2}|\eta_l| 
   \log \left(\sqrt{\frac{\alpha }{2K}} |{\cal X}_{S_l}| \right)
\right]
\notag\\
&\leq \sqrt{\frac{2K}{\alpha}}
\left[
   \phi_{\max} \sum_{l=1}^{L_0} |\xi_l| 
 + \left(\sum_{l=1}^{L_1}|\mu_l|\right) 
  \log \left(\sqrt{\frac{\alpha }{2K}}|\underline{\cal X}|\right)
\right.
\notag\\
&\quad \left.
  + \left(\sum_{l=1}^{L_2}|\eta_l|\right) 
  \log \left(\sqrt{\frac{\alpha }{2K}}|\underline{\cal X}|\right)
  \right]
\notag\\
& \leq {K}^{\prime}
\sqrt{\frac{2K}{\alpha}} 
\log \left(\sqrt{\frac{\alpha }{2K}}{{\rm e}|\underline{{\cal X}}|}\right).
\label{eqn:Ieqb}  
\end{align}

Step (a) follows from the definition of $\phi_{\max}$, 
(\ref{eqn:Ieqa}), and Lemma 2.7 in \cite{ck}. 
From (\ref{eqn:ZxCc}) and (\ref{eqn:Ieqb}), we have 
$$
\tilde{\Psi}_{\max}\leq 
\Psi_{\max}^{(\alpha)} \leq \tilde{\Psi}_{\max}+
{K}^{\prime}
\sqrt{\frac{2K}{\alpha}} 
\log \left(\sqrt{\frac{\alpha }{2K}}{{\rm e}|\underline{{\cal X}}|}\right).
$$
Similarly, we obtain
$$
\tilde{\Psi}_{\min} \geq 
\Psi_{\min}^{(-\alpha)} \geq \tilde{\Psi}_{\min}-
{K}^{\prime}\sqrt{\frac{2K}{\alpha}} 
\log \left(\sqrt{\frac{\alpha }{2K}}{{\rm e}|\underline{{\cal X}}|}\right).
$$
}
We set 
\begin{align*}
   \mu_{\rm sum}&\defeq\sum_{l=1}^{L_1}\mu_l,
   \eta_{\rm sum}\defeq\sum_{l=1}^{L_2}\eta_l, 
\\
\kappa_{\rm sum}&\defeq\sum_{l=1}^{L_3}\kappa_l,
    \nu_{\rm sum}\defeq\sum_{l=1}^{L_4}\nu_l. 
\end{align*}
For $p \in \tilde{\cal P}$ and $\lambda \geq 0$, define
\begin{align*}
& \tilde{\Omega}^{(\lambda)}(p)
\defeq \log {\rm E}_{p}
\left[ \exp \left\{\lambda \omega_{p}(\underline{X}) 
\right\} \right].
\end{align*}
Furthermore, define 
\begin{align}
\tilde{\Omega}_{\max}^{(\lambda)} \defeq \max_{p\in \tilde{\cal P}}
\tilde{\Omega}^{(\lambda)}(p).
\end{align}
For $\lambda \in [-\frac{1}{2 \mu_{\rm sum}}, 
                   \frac{1}{2\eta_{\rm sum}}]$, define 
\begin{align*}
\rho^{(\lambda)} &\defeq
\max_{\scs p \in \tilde{\cal P}:
      \atop{  
      \scs \tilde{\Omega}^{(\lambda)}(p) = \tilde{\Omega}^{(\lambda)}_{\max}
      }
}
{\rm Var}_{p}\left[\omega_{p}(\underline{X})\right].
\end{align*}
Furthermore, set
\begin{align*}
\rho^{(+)} &\defeq \max_{\lambda \in [0, \frac{1}{2\eta_{\rm sum}}]}
\rho^{(\lambda)}, 
\rho^{(-)} \defeq \max_{\lambda \in [-\frac{1}{2\mu_{\rm sum}},0]}
\rho^{(\lambda)}. 
\end{align*}
Note that the quantity $\rho^{(+)}$ depends on $\eta_{\rm sum}$
and the quantity $\rho^{({-})}$ depends on $\mu_{\rm sum}$.
Our main result is given in the following proposition. 
\begin{pro} \label{pro:MainPro}
For any $\alpha$ satisfying 
$\alpha > 2\eta_{\rm sum} +\nu_{\rm sum}$, we have 
\begin{align}
0\leq \Psi^{(\alpha)}_{\max}-\tilde{\Psi}_{\max}\leq 
\frac{1}{\alpha-\nu_{\rm sum}}
\left[ \frac{\rho^{(+)}} {2}+\frac{c^{(+)}}{\alpha-\nu_{\rm sum}}\right],
\label{eqn:MainIeq}
\end{align}
where $c^{(+)}=c^{(+)}(\eta_{\rm sum})$ is a suitable positive 
constant depending on $\eta_{\rm sum}$. Furthermore, for any 
$\alpha$ satisfying 
$\alpha > 2\mu_{\rm sum} + \kappa_{\rm sum}$, we have 
\begin{align}
0\leq \tilde{\Psi}_{\min}-\Psi^{(-\alpha)}_{\min} \leq 
\frac{1}{\alpha-\kappa_{\rm sum}}
\left[\frac{\rho^{(-)}}{2} + \frac{ c^{(-)} }{\alpha-\kappa_{\rm sum}}\right],
\label{eqn:MainIeqX}
\end{align}
where $c^{(-)}=c^{(-)}(\mu_{\rm sum})$ is a suitable positive constant 
depending on $\mu_{\rm sum}$.
\end{pro}

Proof of this proposition will be given in the next section. 
We can see from the above proposition 
that the two bound (\ref{eqn:MainIeq}) and (\ref{eqn:MainIeqX}) 
in Propostion \ref{pro:MainPro}, respectively, provide 
significant improvements from the bounds (\ref{eqn:ProOneIeq}) 
and (\ref{eqn:ProOneIeqX}) in Proposition \ref{pro:ProOne}.

We next consider an application of Propostion \ref{pro:MainPro} 
to the case discussed in Examples 
\ref{ex:ExOne} and \ref{ex:ExTwo}. As stated in Examples \ref{ex:ExOne} 
and \ref{ex:ExTwo}, $\eta_{\rm sum}=\eta_1+\eta_2=2\bar{\xi}$ and 
$\kappa_{\rm sum}=\kappa_1=\bar{\xi}$. 
Set
\begin{align*}
&{\Delta}_{\sf WZ}^{(\xi,\alpha)}(p_{XY}) 
   \defeq {R}_{\sf WZ}^{(\xi)}(p_{XY})
         -{R}_{\sf WZ}^{(\xi,\alpha)}(p_{XY})\geq 0,
\\
&\Delta_{\sf WZ}^{(\alpha)}(p_{XY})\defeq \max_{ \xi \in [0,1]}
 \Delta_{\sf WZ}^{(\xi,\alpha)}(p_{XY}).
\end{align*}
Here we note that $\rho^{(-)}$ and $c^{(-)}$ depend on 
the value of $\xi \in [0,1]$. 
Hence we write $\rho^{(-)}=\rho^{(-)}(\xi)$ 
           and $   c^{(-)}=   c^{(-)}(\xi)$ 
when we wish to express that those are the functions of $\xi$. 
Applying Proposition \ref{pro:MainPro} to the example of Wyner-Ziv source 
coding problem, we have the following result.

\begin{pro}\label{pro:proWZ} 
For any $\xi\in [0,1]$ and any $\alpha$ satisfying 
$\alpha>5\bar{\xi}$, we have 
\begin{align*}
0\leq &   {\Delta}_{\sf WZ}^{(\xi,\alpha)}(p_{XY})
 \leq \frac{1}{\alpha-\bar{\xi}}
          \left[
           \frac{\rho^{(-)}({\xi})}{2} 
         + \frac{   c^{(-)}({\xi})}{\alpha-\bar{\xi}}\right].
\end{align*}
Specifically, for any $\alpha$ satisfying 
$\alpha>5$, we have 
\begin{align*}
0\leq &  {\Delta}_{\sf WZ}^{(\alpha)}(p_{XY})
 \leq \frac{1}{\alpha-1}
          \left[
           \frac{\rho_{\max}^{(-)}}{2} 
         + \frac{   c_{\max}^{(-)}}{\alpha-1} \right],
\end{align*}
where 
$$
\rho_{\max}^{(-)}=\max_{\xi \in [0,1]}\rho^{(-)}(\xi),
   c_{\max}^{(-)}=\max_{\xi \in [0,1]}c^{(-)}(\xi).
$$
\end{pro} 

Let $\varepsilon\in(0,1)$ and for fixed source block length $n$, 
let ${\cal R}_{\rm WZ}(n, \varepsilon| p_{XY})$ 
be the $(n,\varepsilon)$-rate distortion region consisting 
of a pair of compression rate $R$ and distortion level $D$ 
such that the decoder fails to obtain the sources within 
distortion level $D$ with a probability not exceeding 
$\varepsilon$. Formal definition of ${\cal R}_{\rm WZ}(n,\varepsilon| p_{XY})$ 
is found in \cite{OhEnt18wz}. The above theorem together with the result of 
Tyagi and Watanabe \cite{TyagiWatanabeArXiv18} yields 
a new strong converse outer bound. To describe this result
for ${\cal R}\subseteq \mathbb{R}_{+}^2$, we set
\begin{align*}
&{\cal R}-\nu(1,1)\defeq 
\{(a-\nu,b-\nu)\in \mathbb{R}_{+}^2: (a,b)\in{\cal R}\}.
\end{align*}
According to Tyagi and Watanabe \cite{TyagiWatanabeArXiv18}, 
we have the following theorem.
\begin{Th}[Tyagi and Watanabe \cite{TyagiWatanabeArXiv18}]
\label{Th:mainWZth}
For any $\alpha>0$,
\begin{align}
& {\cal R}_{\sf WZ}(n, \varepsilon| p_{XY}) \subseteq {\cal R}_{\sf WZ}(p_{XY})
\notag\\
&\quad -\Biggl\{{\Delta}_{\sf WZ}^{(\alpha)}(p_{XY})
  +\frac{\alpha}{n}\log \frac{1}{1-\varepsilon}\Biggr\}(1,1).
\label{eqn:Zdddcc}
\end{align}
\end{Th}

From Theorem \ref{Th:mainWZth} and Proposition \ref{pro:proWZ}, 
we have the following:  
\begin{Th}For any $\alpha$ satisfying $\alpha>5$, we have 
\begin{align}
&          {\cal R}_{\sf WZ}(n,\varepsilon| p_{XY})
 \subseteq {\cal R}_{\sf WZ}(p_{XY})
           -\upsilon_n(\varepsilon,\alpha)(1,1),
\end{align}
where
\begin{align}
&\upsilon_n(\varepsilon,\alpha)
\defeq 
\frac{1}{\alpha-1}\left[\frac{\rho_{\max}^{(-)}}{2}
        + \frac{ c_{\max}^{(-)} }{\alpha-1}\right]
+\frac{\alpha}{n}\log \frac{1}{1-\varepsilon}.
\label{eqn:TWouterbound}
\end{align}
\end{Th}
In (\ref{eqn:TWouterbound}), we choose 
$
\ts
\alpha=\alpha_n(\varepsilon) 
=\sqrt{ 
     \frac{\rho_{\max}^{(-)}n}{2\log\frac{1}{1-\varepsilon}}}+1.
$
For this choice of $\alpha_n(\varepsilon)$, the quantity
$\upsilon_n(\varepsilon)=\upsilon_n(\varepsilon,$$\alpha_n(\varepsilon))$ 
becomes the following:
$$
\upsilon_n(\varepsilon)=
\sqrt{ \frac{2\rho_{\max}^{(-)}}{n}\log \frac{1}{1-\varepsilon}}
 +\frac{1}{n}\left[\frac{2c_{\max}^{(-)}}{\rho_{\max}^{(-)} }+1 \right]
 \log \frac{1}{1-\varepsilon}. 
$$
The quantity $\upsilon_n(\varepsilon)$ indicates a gap of the outer 
bound of ${\cal R}_{\sf WZ}(n, \varepsilon| p_{XY})$ from 
${\cal R}_{\sf WZ}(p_{XY})$. This gap is tighter than 
the similar gap $\upsilon_n^{\prime}(\varepsilon)$ 
given by 
$$
 \upsilon_n^{\prime}(\varepsilon)
=\sqrt{\frac{c}{n}\log\frac{5}{1-\varepsilon}}+
\frac{2}{n}\log\frac{5}{1-\varepsilon},
$$
where $c$ is some positive constant not depending on $(n,\varepsilon)$.
The above $\upsilon_n^{\prime}(\varepsilon)$ was obtained by 
the author \cite{OhEnt18wz} in a different method based on the theory 
of information spectrums \cite{han}.

\section{Proof of the Main Result}

For $\theta, \alpha \geq 0$, and for $q \in {\cal P}^{*}$, define 
\begin{align*}
 \omega^{(\alpha)}_{q}(\underline{x})&
\defeq \omega_{q}(\underline{x})
-\alpha \log \frac{q(\underline{x})}{p^{(q)}(\underline{x})},
\\
 \Omega^{(\theta,\alpha)}(q) &
\defeq \log {\rm E}_{q}
\left[\exp\left\{\theta
\omega^{(\alpha)}_{q}(\underline{X})\right\}\right].
\end{align*}
We can show that the functions we have definded so far satisfy several 
properties shown below. 
\begin{pr}\label{pr:pro1}  
$\quad$
\begin{itemize}

\item[a)] 
For fixed positive $\alpha>0$, a sufficient condition 
for $\Omega^{(\theta,\alpha)}$ to exist is 
$
\theta \in \left[0,\frac{1}{\alpha+\eta_{\rm sum}}\right].  
$
\item[b)] 
For $q \in {\cal P}^{*}$, define a probability distribution 
$q^{(\theta;\alpha)}$ by
\beqno 
& &q^{(\theta;\alpha)}(\underline{x})
\defeq
\frac{q(\underline{x})
\exp\left\{\theta {\omega}^{(\alpha)}_{q}(\underline{x})
\right\}}
{{\rm E}_{q}
\left[\exp\left\{\theta
{\omega}^{(\alpha)}_{q}(\underline{X})\right\}\right]}.
\eeqno
For $p \in \tilde{\cal P}^{*}$, define 
a probability distribution $p^{(\lambda)}$ by
\beqno 
& &p^{(\lambda)}(\underline{x})
\defeq
\frac{p(\underline{x})
\exp\left\{\lambda {\omega}_p(\underline{x})\right\}}
{{\rm E}_p \left[\exp\left\{\lambda 
\omega_p(\underline{X})\right\}\right]}.
\eeqno
Then, we have
\begin{align}
  \frac{\rm {d}}{{\rm {d}}\theta}
{\Omega}^{(\theta,\alpha)}(q)
&={\rm E}_{q^{(\theta;\alpha)}}
\left[\omega^{(\alpha)}_{q}(\underline{X})\right],
\label{eqn:aabb0}\\
 \frac{\rm {d}^2}{{\rm {d}}\theta^2} 
{\Omega}^{(\theta,\alpha)}(q)
&={\rm Var}_{q^{(\theta;\alpha)}}
\left[{\omega}^{(\alpha)}_{q}(\underline{X})\right],
\label{eqn:aabb1}\\
 \frac{\rm d}{{\rm d}\lambda}
\tilde{\Omega}^{(\lambda)}(p)
&={\rm E}_{p^{(\lambda)}} \left[\omega_{p}(\underline{X})\right],
\label{eqn:aabb2}\\
 \frac{\rm d^2}{{\rm d}\lambda^2} 
\tilde{\Omega}^{(\lambda)}(p)
&={\rm Var}_{{p}^{(\lambda)}} \left[\omega_{p}(\underline{X})\right].
\label{eqn:aabb3}
\end{align}
Specifically, we have 
\begin{align}
\biggl( \frac{\rm {d}}{{\rm {d}}\theta}
{\Omega}^{(\theta,\alpha)}(q)\biggr)_{\theta=0}
&={\rm E}_q \left[\omega^{(\alpha)}_{q}(\underline{X})\right]
={\Psi}^{(\alpha)}(q),
\label{eqn:aabbx0}\\
\biggl(\frac{\rm d}{{\rm d}\lambda}\tilde{\Omega}^{(\lambda)}(p)
\biggr)_{\lambda=0}
& ={\rm E}_{p}\left[\omega_{p}(\underline{X})\right]=\tilde{\Psi}(p).
\label{eqn:aabbx2}
\end{align}
For fixed $\alpha>0$, a sufficient condition for the 
three times derivative of $\Omega^{(\theta,\alpha)}$ 
to exist is 
$
\theta \in \left[0,\frac{1}{2(\alpha+\eta_{\rm sum})}\right]. 
$
Furthermore, a sufficient condition for the three times 
derivative of $\Omega^{(\lambda)}$ to exist is 
$\lambda \in [-\frac{1}{2\mu_{\rm sum}}, 
     \frac{1}{2\eta_{\rm sum}}]$.
\item[c)] 
Let $c^{(+)}=c^{(+)}(\eta_{\rm sum})$ be some positive constant 
depending on $\eta_{\rm sum}$. 
Then, for any  $\lambda \in [0,\frac{1}{2\eta_{\rm sum}}]$, we have
\begin{align}
&\tilde{\Omega}_{\max}^{(\lambda)}
\leq \lambda \tilde{\Psi}_{\max} 
+ {\lambda^2}\left[ \frac{{\rho}^{(+)}}{2} + \lambda c^{(+)} \right]. 
\label{eqn:Sdxxz}
\end{align}
\newcommand{\Zasaa}{
From (\ref{eqn:AiSs2}) and (\ref{eqn:Sdxxz}), we have that 
for any $ \alpha >\nu_{\rm sum} $ and any $q\in {\cal P}^{*}$, 
\beq
\Psi^{(\alpha)}(q) \leq 
\tilde{\Psi}_{\max}+\frac{1}{\alpha-\nu_{\rm sum}}
\left[\frac{\rho}{2}+\frac{c}{\alpha-\nu_{\rm sum}}\right].
\label{eqn:AszzcE}
\eeq
}
\item[d)] 
For any $q \in {\cal P}^{*}$, 
$p^{(q)}=\varphi(q)\in \tilde{\cal P}$, any $\alpha>\nu_{\rm sum}$, 
and any $\theta \in [0,\frac{1}{\alpha+\kappa_{\rm sum}}]$,
we have 
\begin{align}
&{\Omega}^{(\theta,\alpha)}(q)
\leq \theta(\alpha-\nu_{\rm sum})
\tilde{\Omega}^{(\frac{1}{\alpha-\nu_{\rm sum}})}(p^{(q)}).
\label{eqn:AiSs0}
\end{align}
From (\ref{eqn:AiSs0}), we have
\begin{align}
&\frac{\Omega^{(\theta,\alpha)}(q)}{\theta}
\leq (\alpha-\nu_{\rm sum})
\tilde{\Omega}^{(\frac{1}{\alpha-\nu_{\rm sum}})}(p^{(q)})
\label{eqn:AiSs1}
\end{align}
for $\theta \in (0,\frac{1}{\alpha+\kappa_{\rm sum}}]$.
By letting $\theta \to 0$ in (\ref{eqn:AiSs1}), and taking 
(\ref{eqn:aabbx0}) into account, we have that for any 
$\alpha >\nu_{\rm sum}$ and any $q \in {\cal P}^{*}$, 
$p^{(q)}\in \tilde{\cal P}$, 
\begin{align}
&\Psi^{(\alpha)}(q)
\leq (\alpha - \nu_{\rm sum})
\tilde{\Omega}^{(\frac{1}{\alpha-\nu_{\rm sum}})}(p^{(q)}).
\label{eqn:AiSs2}
\end{align}
\end{itemize}
\end{pr}
\newcommand{\ApdaAAAb}{}
\newcommand{\ApdaAAAbZZ}{}
\begin{pr}\label{pr:pro1b}  
$\quad$
\begin{itemize}
\item[a)] For fixed positive $\alpha>0$, a sufficient condition 
for $\Omega^{(\theta,-\alpha)}$ to exist is 
$
\theta \in \left[\frac{-1}{\alpha+\mu_{\rm sum}},0 \right].
$
\item[b)] 
For fixed positive $\alpha>0$, a sufficient condition 
for the three times derivative of $\Omega^{(\theta,-\alpha)}$ 
to exist is 
$
\theta \in \left[\frac{-1}{2(\alpha+\mu_{\rm sum})},0 \right].  
$
\item[c)] 
Let $c^{(-)}=c^{(-)}(\mu_{\rm sum})$ be some positive constant 
depending on $\mu_{\rm sum}$. 
Then, for any  $\lambda \in [0,\frac{1}{2\mu_{\rm sum}}]$, we have
\begin{align}
&\tilde{\Omega}_{\max}^{(-\lambda)}
\leq -\lambda \tilde{\Psi}_{\min} 
+  {\lambda^2}\left[ \frac{{\rho}^{(-)}}{2} + \lambda c^{(-)}\right]. 
\label{eqn:Sdxxaz}
\end{align}
\item[d)]
For any $q \in {\cal P}^{*}$, 
$p^{(q)}=\varphi(q)\in \tilde{\cal P}$, any $\alpha>\kappa_{\rm sum}$, 
and any $\theta \in [\frac{-1}{\alpha+\nu_{\rm sum}},0]$,
we have 
\begin{align}
&{\Omega}^{(\theta,-\alpha)}(q)
\leq -\theta(\alpha-\kappa_{\rm sum})
\tilde{\Omega}^{(\frac{-1}{\alpha-\kappa_{\rm sum}})}(p^{(q)}).
\label{eqn:AiSs0z}
\end{align}
From (\ref{eqn:AiSs0z}), we have
\begin{align}
&\frac{\Omega^{(\theta,-\alpha)}(q)}{\theta}
\geq -(\alpha-\kappa_{\rm sum})
\tilde{\Omega}^{(\frac{-1}{\alpha-\kappa_{\rm sum}})}(p^{(q)})
\label{eqn:AiSs1z}
\end{align}
for $\theta \in [\frac{-1}{\alpha+\nu_{\rm sum}},0)$.
By letting $\theta \to 0$ in (\ref{eqn:AiSs1z}), and taking 
(\ref{eqn:aabbx0}) into account, we have that for any 
$\alpha > \kappa_{\rm sum}$ and any $q \in {\cal P}^{*}$, 
$p^{(q)}\in \tilde{\cal P}$, 
\begin{align}
&\Psi^{(-\alpha)}(q)
\geq -(\alpha -\kappa_{\rm sum})
\tilde{\Omega}^{(\frac{-1}{\alpha-\kappa_{\rm sum}})}(p^{(q)}).
\label{eqn:AiSs2z}
\end{align}
\end{itemize}
\newcommand{\Zasdd}{
From (\ref{eqn:AiSs2z}) and (\ref{eqn:Sdxxaz}), we have that 
for any $\alpha >\kappa_{\rm sum} $ and any $q\in {\cal P}^{*}$, 
\beq
\Psi^{(-\alpha)}(q) \geq 
\tilde{\Psi}_{\min}- \frac{1}{\alpha-\kappa_{\rm sum}}
\left[\frac{\rho}{2}+\frac{c^{\prime}}{\alpha-\kappa_{\rm sum}}\right].
\label{eqn:AszzcEx}
\eeq
}
\end{pr}

Proofs of Properties \ref{pr:pro1} and \ref{pr:pro1b} part a)-c) are given 
in Appendix \ref{sub:ApdaAAC}. 
Proofs of the equalities in Property \ref{pr:pro1} part b) 
are also given in Appendix \ref{sub:ApdaAAC}. 
Proofs 
of  the inequality (\ref{eqn:AiSs0})  in Property \ref{pr:pro1}  part d)
and the inequality (\ref{eqn:AiSs0z}) in Property \ref{pr:pro1b} part d)
are given in Appendix \ref{sub:ApdaAACa}. 

{\it Proof of Propositon \ref{pro:MainPro}: }
We first prove (\ref{eqn:MainIeq}). Fix $q \in {\cal P}^{*}$ arbitrary.
For $\alpha > \nu_{\rm sum}$, we set
$\alpha=\lambda^{-1} + \nu_{\rm sum}.$
Then the condition $\alpha > 2\eta_{\rm sum} + \nu_{\rm sum}$ is 
equivalent to $ \lambda \in (0,\frac{1}{2\eta_{\rm sum}}]$.
When $\alpha> 2\eta_{\rm sum} + \nu_{\rm sum},$
we have the following chain of inequalities  
\begin{align} 
\Psi^{(\alpha)}(q) 
&\MLeq{a} \frac{1}{\lambda}\tilde{\Omega}(p^{(q)})
\leq  \frac{\tilde{\Omega}_{\max}^{(\lambda)}}{\lambda}
\notag\\
&\MLeq{b}
\tilde{\Psi}_{\max}
+{\lambda}\left[ \frac{{\rho}^{(+)}}{2}+\lambda c^{(+)} \right] 
\notag\\
&=\tilde{\Psi}_{\max}+ 
\frac{1}{\alpha-\nu_{\rm sum}}\left[ \frac{\rho^{(+)}} {2}
+\frac{c^{(+)}}{\alpha-\nu_{\rm sum}}\right].
\label{eqn:AxxP}
\end{align} 
Step (a) follows from (\ref{eqn:AiSs2}) in Property \ref{pr:pro1} part d)
and the choice $\lambda=(\alpha-\nu_{\rm sum})^{-1}$ of $\lambda$.
Step (b) follows from (\ref{eqn:Sdxxz}) in Property \ref{pr:pro1} part c).
Since (\ref{eqn:AxxP}) holds for any $q\in {\cal P}^{*}$, we have 
(\ref{eqn:MainIeq}) in Proposition \ref{pro:MainPro}.
We next prove (\ref{eqn:MainIeqX}). Fix $q \in {\cal P}^{*}$ arbitrary.
For $\alpha > \kappa_{\rm sum}$, we set
$\alpha=\lambda^{-1} + \kappa_{\rm sum}.$
Then the condition $\alpha > 2\mu_{\rm sum} + \kappa_{\rm sum}$ is 
equivalent to $\lambda \in (0, \frac{1}{2\mu_{\rm sum}}]$.
When $\alpha>2\mu_{\rm sum} + \kappa_{\rm sum},$
we have the following chain of inequalities:  
%
%
%
%
%
%
%
%
%
%
\begin{align} 
\Psi^{(-\alpha)}(q) 
&\MGeq{a} \frac{1}{-\lambda}\tilde{\Omega}^{(-\lambda)}(p^{(q)})
\geq \frac{\tilde{\Omega}_{\max}^{(-\lambda)}}{-\lambda}
\notag\\
&\MGeq{b}
\tilde{\Psi}_{\min}-{\lambda}\left[ \frac{{\rho}^{(-)}}{2}
+\lambda c^{(-)} \right] 
\notag\\
&=\tilde{\Psi}_{\min}
 -\frac{1}{\alpha-\kappa_{\rm sum}}\left[ \frac{\rho^{(-)}} {2}
 +\frac{c^{(-)}}{\alpha-\kappa_{\rm sum}}\right].
\label{eqn:AxxPz}
\end{align}
Step (a) follows from (\ref{eqn:AiSs2z}) in Property \ref{pr:pro1b} part d) 
and the choice $\lambda=(\alpha-\kappa_{\rm sum})^{-1}$ of $\lambda$.
Step (b) follows from (\ref{eqn:Sdxxaz}) in Property \ref{pr:pro1b} part c).
Since (\ref{eqn:AxxPz}) holds for any $q\in {\cal P}^{*}$, we have 
(\ref{eqn:MainIeqX}) in Proposition \ref{pro:MainPro}.
\hfill\IEEEQED


\newcommand{\ApdaAAC}{


\subsection{
Proofs of Properties \ref{pr:pro1} and \ref{pr:pro1b}, Parts a)-c)
}\label{sub:ApdaAAC}

In this appendix we prove Properties \ref{pr:pro1} and \ref{pr:pro1b}
parts a), b) and c). We first prove Properties \ref{pr:pro1} 
and \ref{pr:pro1b} part a).

{\it Proof of Property \ref{pr:pro1} parts a) and b):} 
We first prove the part a). 
For $\alpha>0$, we have the following form of 
$\exp\{\Omega^{(\theta,\alpha)}(q)\}$:
\begin{align*}
&\exp\{\Omega^{(\theta,\alpha)}(q)\}=\sum_{\scs 
\underline{x} \in \underline {\cal X}}
q^{1-\theta\alpha}(\underline{x})
p^{\theta\alpha}(\underline{x})
\left( \prod_{l=1}^{L_0}{\rm e}^{\xi_l \phi_l(\underline{x}) }\right)
\notag\\
&\quad \times
\left(\prod_{l=1}^{L_1} 
\left[ q_{{X}_{S_l}}(x_{S_l})\right]^{\theta\mu_l} \right)
\left(\prod_{l=1}^{L_2}  
\left[q_{{X}_{T_l}}(x_{T_l}) \right]^{-\theta\eta_l}\right) 
\notag\\
&= \sum_{\scs 
 \underline{x} \in \underline{\cal X}}
 p^{\theta\alpha}(\underline{x})
\left(\prod_{l=1}^{L_0}{\rm e}^{\xi_l \phi_l(\underline{x})}\right)
\left(\prod_{l=1}^{L_1} 
q^{\mu_l{\theta}}_{{X}_{S_l}}(x_{S_l})\right)
\notag\\
&\quad \times
\left(\prod_{l=1}^{L_2}  
q^{\eta_l}(\underline{x})
\right)^{\frac{1-(\alpha+\eta_{\rm sum})\theta}{\eta_{\rm sum}}}
\left(\prod_{l=1}^{L_2}  
q^{\eta_l\theta}_{X_{T_l^c}|{X}_{T_l}}(x_{T^c_l}|x_{T_l})
\right)
\end{align*}
from which we can see that 
\begin{align}
&\theta \geq 0, 
  \frac{1-(\alpha+\eta_{\rm sum})\theta}{\eta_{\rm sum}}
  \geq 0 \Leftrightarrow 
  \theta \in \left[0,\frac{1}{\alpha+\eta_{\rm sum}}\right] 
\label{eqn:CondA}
\end{align}
is a sufficient condition for $\exp\{\Omega^{(\theta,\alpha)}(q)\}$ 
to be bounded and strictly positive. 
Hence, (\ref{eqn:CondA}) is a sufficient condition 
for $\Omega^{(\theta,\alpha)}(q)$ to exist. 
We have the following form of 
$\exp\left\{ \tilde{\Omega}^{(\lambda)}(p)\right\}$:
\begin{align*}
&\exp\left\{\tilde{\Omega}^{(\lambda)}(p) \right\}
=\sum_{\scs 
\underline{x} \in \underline {\cal X}}
p(\underline{x})
\left( \prod_{l=1}^{L_0}{\rm e}^{\xi_l \phi_l(\underline{x}) }\right)
\notag\\
&\quad \times
\left(\prod_{l=1}^{L_1} 
\left[ p_{{X}_{S_l}}(x_{S_l})\right]^{\lambda\mu_l} \right)
\left(\prod_{l=1}^{L_2}  
\left[p_{{X}_{T_l}}(x_{T_l}) \right]^{-\lambda\eta_l}\right) 
\notag\\
&= \sum_{\scs 
 \underline{x} \in \underline{\cal X}}
\left(\prod_{l=1}^{L_0}{\rm e}^{\xi_l \phi_l(\underline{x})}\right)
\left(\prod_{l=1}^{L_1} 
p^{\mu_l{\lambda}}_{{X}_{S_l}}(x_{S_l})\right)
\notag\\
&\quad \times
\left(\prod_{l=1}^{L_2}  
p^{\eta_l}(\underline{x})
\right)^{\frac{1-\eta_{\rm sum}\lambda}{\eta_{\rm sum}}}
\left(\prod_{l=1}^{L_2}  
p^{\eta_l\lambda}_{X_{T_l^c}|{X}_{T_l}}(x_{T^c_l}|x_{T_l})
\right)
\notag\\
&= \sum_{\scs 
 \underline{x} \in \underline{\cal X}}
\left(\prod_{l=1}^{L_0}{\rm e}^{\xi_l \phi_l(\underline{x})}\right)
\left(\prod_{l=1}^{L_2} 
p^{-\eta_l{\lambda}}_{{X}_{T_l}}(x_{T_l})\right)
\notag\\
&\quad \times
\left(\prod_{l=1}^{L_1}  
p^{\mu_l}(\underline{x})
\right)^{\frac{1+\mu_{\rm sum}\lambda}{\mu_{\rm sum}}}
\left(\prod_{l=1}^{L_1}  
p^{-\mu_l\lambda}_{X_{T_l^c}|{X}_{T_l}}(x_{T^c_l}|x_{T_l})
\right)
\end{align*}
from which we can see that 
\begin{align}
& \lambda\geq 0, {\frac{1-\eta_{\rm sum}\lambda}{\eta_{\rm sum}}}\geq 0 
\mbox{ or }\lambda \leq 0, 
{\frac{1+\mu_{\rm sum}\lambda}{\mu_{\rm sum}}}\geq 0
\notag\\
&\Leftrightarrow \lambda 
\in \left[-\frac{1}{\mu_{\rm sum}}, \frac{1}{\eta_{\rm sum}}\right]
\label{eqn:CondAa} 
\end{align}
is a sufficient condition for $\exp\{\tilde{\Omega}^{(\lambda)}(p)\}$ 
to be bounded and strictly positive. Hence, (\ref{eqn:CondAa}) is 
a sufficient condition for $\tilde{\Omega}^{(\lambda)}(p)$ 
to exist. 
\hfill\IEEEQED

{\it Proof of Property \ref{pr:pro1b} part a):} 
For $\alpha>0$, we have the following form of 
$\exp\{\Omega^{(\theta,-\alpha)}(q)\}$:
\begin{align*}
&\exp\left\{ \Omega^{(\theta,-\alpha)}(q)\right\}
=\sum_{\scs 
\underline{x} \in \underline {\cal X}}
q^{1+\theta\alpha}(\underline{x})
p^{-\theta\alpha}(\underline{x})
\left( \prod_{l=1}^{L_0}{\rm e}^{\xi_l \phi_l(\underline{x}) }\right)
\notag\\
&\quad \times
\left(\prod_{l=1}^{L_2} 
\left[ q_{{X}_{T_l}}(x_{T_l})\right]^{-\theta \eta_l} \right)
\left(\prod_{l=1}^{L_1}  
\left[q_{{X}_{S_l}}(x_{S_l}) \right]^{\theta \mu_l}\right) 
\notag\\
&= \sum_{\scs 
 \underline{x} \in \underline{\cal X}}
 p^{-\theta\alpha}(\underline{x})
\left(\prod_{l=1}^{L_0}{\rm e}^{\xi_l \phi_l(\underline{x})}\right)
\left(\prod_{l=1}^{L_2} 
q^{-\mu_l\theta}_{{X}_{T_l}}(x_{T_l})\right)
\notag\\
&\quad \times
\left(\prod_{l=1}^{L_1}  
q^{\mu_l}(\underline{x})
\right)^{\frac{1+(\alpha+\mu_{\rm sum})\theta}{\mu_{\rm sum}}}
\left(\prod_{l=1}^{L_1}  
q^{-\mu_l\theta}_{X_{S_l^c}|{X}_{S_l}}(x_{S^c_l}|x_{S_l})
\right)
\end{align*}
from which we can see that 
\begin{align}
& \theta \leq 0, 
  \frac{1+(\alpha+\mu_{\rm sum})\theta}{\mu_{\rm sum}}
  \geq 0 
  \Leftrightarrow \theta \in \left[\frac{-1}{\alpha+\mu_{\rm sum}},0 \right]  
\label{eqn:CondBb}
\end{align}
is a sufficient condition for $\exp\{\Omega^{(\theta,-\alpha)}(q)\}$ 
to be bounded and strictly positive. Hence, (\ref{eqn:CondBb}) is a 
sufficient condition for $\Omega^{(\theta,-\alpha)}(q)$ to exist. 
\hfill\IEEEQED

We next prove Properties \ref{pr:pro1} and \ref{pr:pro1b} part b).
For simplicity of notations, set
\begin{align*}
& \omega^{(\alpha)}_{q}(\underline{x}) \defeq \varsigma(\underline{x}),
\Omega^{(\theta, \alpha)}(q)\defeq \zeta(\theta),
\\
& \omega_{q}(\underline{x}) \defeq \tilde{\varsigma}(\underline{x}),
\tilde{\Omega}^{(\lambda)}(q) \defeq \tilde{\zeta}(\lambda).
\end{align*}
Then we have 
\begin{align}
& \Omega^{(\theta,\alpha)}(q)=\zeta(\theta)=\log
\left[
\sum_{\underline{x}\in \underline{\cal X} }q(\underline{x})
{\rm e}^{\theta{\varsigma}(\underline{x})}
\right].
\label{eqn:SdfV}
\\
& \tilde{\Omega}^{(\lambda)}(q)=\tilde{\zeta}(\theta)=\log
\left[\sum_{\underline{x}\in \underline{\cal X} }q(\underline{x})
{\rm e}^{\theta \tilde{\varsigma}(\underline{x})}\right].
\label{eqn:SdfVzzz}
\end{align}
For each $\underline{x}\in {\cal X}$,
the quantities 
$
q^{(\theta;\alpha)}(\underline{x})$
and 
$q^{(\lambda)}(\underline{x})$ have the following forms:
\begin{align}
 q^{(\theta;\alpha)}(\underline{x})
 &={\rm e}^{-\zeta(\theta)}q(\underline{x})
 {\rm e}^{\theta{\varsigma}(\underline{x})},
\label{eqn:aazP011}
\\
 q^{(\lambda)}(\underline{x})
  &={\rm e}^{-\tilde{\zeta}(\lambda)}q(\underline{x})
  {\rm e}^{\lambda \tilde{\varsigma}(\underline{x})}.
 \label{eqn:aazP011z}
\end{align}
By simple computations we have 
\begin{align}
& \zeta^{\prime}(\theta)=
{\rm e}^{-\xi(\theta)}
\sum_{\underline{x}}q(\underline{x})
\varsigma(\underline{x})
{\rm e}^{\theta {\varsigma}(\underline{x})}
=\sum_{\underline{x}}q^{(\theta;\alpha)}(\underline{x})
\varsigma(\underline{x}),
\label{eqn:az011}\\
&\zeta^{\prime\prime}(\theta)
=\sum_{\underline{x}\in \underline{\cal X}}
q^{(\theta;\alpha)}(\underline{x})\varsigma^2(\underline{x})
-\left[
\sum_{\underline{x}\in \underline{\cal X}} 
q^{(\theta;\alpha)}(\underline{x})
{\varsigma}(\underline{x})\right]^2,
\label{eqn:aaz011}
\\
&\zeta^{\prime\prime\prime}(\theta)=
\sum_{\underline{x}\in \underline{\cal X}}
q^{(\theta; \alpha)}(\underline{x})\varsigma^3(\underline{x})
\nonumber\\
&
\quad-3\left[
\sum_{\underline{x}\in \underline{\cal X}} 
q^{(\theta;\alpha)}(\underline{x})
{\varsigma}^2(\underline{x})\right]
\left[
\sum_{\underline{x}\in \underline{\cal X}} 
q^{(\theta;\alpha)}(\underline{x})
{\varsigma}(\underline{x})\right]
\nonumber\\
&\quad 
+2\left[
\sum_{\underline{x}\in \underline{\cal X}} 
q^{(\theta;\alpha)}(\underline{x})
{\varsigma}(\underline{x})\right]^3.
\label{eqn:aaaz011}
\end{align}
By simple computations we have 
\begin{align}
& \tilde{\zeta}^{\prime}(\lambda)=
{\rm e}^{-\tilde{\zeta}(\lambda)}
\sum_{\underline{x}}q(\underline{x})
\tilde{\varsigma}(\underline{x})
{\rm e}^{\lambda \tilde{\varsigma}(\underline{x})}
=\sum_{\underline{x}}q^{(\lambda)}(\underline{x})
\tilde{\varsigma}(\underline{x})
\label{eqn:az011zz}\\
&\tilde{\zeta}^{\prime\prime}(\lambda)
=\sum_{\underline{x}\in \underline{\cal X}}
q^{(\lambda)}(\underline{x})\tilde{\varsigma}^2(\underline{x})
-\left[
\sum_{\underline{x}\in \underline{\cal X}} 
q^{(\lambda)}(\underline{x})
{\varsigma}(\underline{x})\right]^2,
\label{eqn:aaz011zzz}
\\
&\tilde{\zeta}^{\prime\prime\prime}(\lambda)=
\sum_{\underline{x}\in \underline{\cal X}}
q^{(\lambda)}(\underline{x})\tilde{\varsigma}^3(\underline{x})
\nonumber\\
&
\quad-3\left[
\sum_{\underline{x}\in \underline{\cal X}} 
q^{(\lambda)}(\underline{x})
\tilde{\varsigma}^2(\underline{x})\right]
\left[
\sum_{\underline{x}\in \underline{\cal X}} 
q^{(\lambda)}(\underline{x})
\tilde{\varsigma}(\underline{x})\right]
\nonumber\\
&\quad 
+2\left[
\sum_{\underline{x}\in \underline{\cal X}} 
q^{(\lambda)}(\underline{x})
\tilde{\varsigma}(\underline{x})\right]^3.
\label{eqn:aaaz011zz}
\end{align}

The following lemma is useful to derive sufficient 
conditions for the existances of 
three times derivative of 
$\Omega^{(\theta,\alpha)}(q)$ and 
$\tilde{\Omega}^{(\lambda)}(p)$ to exist.

\begin{lm}\label{lm:lmDev}
A condition for the following quantity 
$$
{\rm e}^{\zeta(2\theta)-2\zeta(\theta)}
=\frac{\exp\{\Omega^{(2\theta,\alpha)}(q)\}}
      {\exp\{2\Omega^{(\theta,\alpha)}(q)\}}
$$
to be bounded is a sufficient condition for 
the three times derivative of 
$\Omega^{(\theta,\alpha)}(q)$ to exist. 
Similarly, a condition for the following quantity 
$$
{\rm e}^{\tilde{\zeta}(2\lambda)-2\tilde{\zeta}(\lambda)}
=\frac{\exp\{\tilde{\Omega}^{(2\lambda)}(p)\}}
      {\exp\{2\tilde{\Omega}^{(\lambda)}(p)\}}
$$
to be bounded is a sufficient condition for 
the three times derivative of 
$\tilde{\Omega}^{(\lambda)}(p)$ to exist.
\end{lm}  

{\it Proof:} We only prove the first claim.
The second claim can be proved by a quite parallel argument. 
We omit the detail. By 
(\ref{eqn:az011}),
(\ref{eqn:aaz011}), 
and (\ref{eqn:aaaz011}), 
we can see that if 
$$
\sum_{\underline{x}\in \underline{\cal X}}
q^{(\lambda)}(\underline{x})|\varsigma^j(\underline{x})|
$$
are bounded for $j=1,2,3$, the three quantities
$\zeta^{\prime}(\theta),$ 
$\zeta^{\prime\prime}(\theta),$ and 
$\zeta^{\prime\prime\prime}(\theta)$ are also bounded. 
We have the following chain of inequalities:
\begin{align}
&\sum_{\underline{x}\in \underline{\cal X}}
q^{(\theta)}(\underline{x})|\varsigma^j(\underline{x})|
\MEq{a}{\rm e}^{-\zeta(\theta)}
\sum_{\underline{x}\in \underline{\cal X}}
q(\underline{x})|\varsigma^j(\underline{x})|
{\rm e}^{\theta{\varsigma}(\underline{x})}
\nonumber\\
&\MLeq{b} 
\left[ \sum_{\underline{x}\in \underline{\cal X}}
q(\underline{x})\varsigma^{2j}(\underline{x})
\right]^{\frac{1}{2}}
{\rm e}^{\frac{1}{2}\zeta(2\theta)-\zeta(\theta)}.
\label{eqn:Sdppi}
\end{align}
Step (a) follows from (\ref{eqn:aazP011}).
Step (b) follows from Cauchy-Schwarz inequality and 
(\ref{eqn:SdfV}). From (\ref{eqn:Sdppi}) and the well-known fact  
$$
\sum_{\underline{x}\in \underline{\cal X}}
q(\underline{x})\varsigma^{2j}(\underline{x})< \infty \mbox{ for }j=1,2,3,
$$
we can see that 
a condition for the following quantity 
$$
{\rm e}^{\zeta(2\theta)-2\zeta(\theta)}
=\frac{\exp\{\Omega^{(2\theta,\alpha)}(q)\}}
      {\exp\{2\Omega^{(\theta,\alpha)}(q)\}}
$$
to be bounded is a sufficient condition for 
the three times derivative of 
$\Omega^{(\theta,\alpha)}(q)$ to exist. 
Similarly, a condition for the following quantity 
$$
{\rm e}^{\tilde{\zeta}(2\lambda)-2\tilde{\zeta}(\lambda)}
=\frac{\exp\{\tilde{\Omega}^{(2\lambda)}(p)\}}
      {\exp\{2\tilde{\Omega}^{(\lambda)}(p)\}}
$$
to be bounded is a sufficient condition for 
the three times derivative of 
$\tilde{\Omega}^{(\lambda)}(p)$ to exist.
\hfill \IEEEQED 

{\it Proof of Property \ref{pr:pro1} part b):} 
By Property \ref{pr:pro1} part a), when
\begin{align}
\ds 
  2\theta \in \left[0,\frac{1}{\alpha+\eta_{\rm sum}}\right] 
  \Leftrightarrow 
  \theta \in \left[0,\frac{1}{2(\alpha+\eta_{\rm sum})}\right],
\label{eqn:CondAb}
\end{align}
$\Omega^{(2\theta,\alpha)}(q)$ and $\Omega^{(\theta,\alpha)}(q)$ 
exist. Hence, by Lemma \ref{lm:lmDev}, (\ref{eqn:CondAb}) is 
a sufficient condition for $\Omega^{(\theta,\alpha)}(q)$ to be 
three times differentiable. By Property \ref{pr:pro1} part a), when
\begin{align}
& 2\lambda 
\in \left[-\frac{1}{\mu_{\rm sum}}, \frac{1}{\eta_{\rm sum}}\right]
\Leftrightarrow  
\lambda 
\in \left[-\frac{1}{2\mu_{\rm sum}}, 
\frac{1}{2\eta_{\rm sum}}\right],
\label{eqn:CondAbzz}
\end{align}
    $\tilde{\Omega}^{(2\lambda)}(p)$ 
and $\tilde{\Omega}^{(\lambda)}(p)$ 
exist. Hence, by Lemma \ref{lm:lmDev}, (\ref{eqn:CondAbzz}) is 
a sufficient condition for $\tilde{\Omega}^{(\lambda)}(p)$ to be 
three times differentiable. 
\hfill\IEEEQED

{\it Proof of Property \ref{pr:pro1b} part b):}
By Property \ref{pr:pro1b} part a), when
\begin{align}
&  2\theta \in \left[\frac{-1}{\alpha+\mu_{\rm sum}},0 \right] 
  \Leftrightarrow 
  \theta \in \left[\frac{-1}{2(\alpha+\mu_{\rm sum})},0 \right],  
  \label{eqn:CondBc}
\end{align}
$\Omega^{(2\theta,-\alpha)}(q)$ and $\Omega^{(\theta,-\alpha)}(q)$ 
exist. Hence, by Lemma \ref{lm:lmDev}, (\ref{eqn:CondBc}) is 
a sufficient condition for $\Omega^{(\theta,-\alpha)}(q)$ to be 
three times differentiable. 
\hfill\IEEEQED


We finally prove Properties \ref{pr:pro1} and  \ref{pr:pro1b} part c). 

{\it Proof of Property \ref{pr:pro1} part c):} 
Fix any $p \in \tilde{\cal P}$. 
By the Taylor expansion of 
$\tilde{\Omega}^{(\lambda)}(p)$
with respect to $\lambda$ around $\lambda=0$,
we have that 
for any $(p, \lambda)\in$
$\tilde{\cal P}\times$ $[0, \frac{1}{2\eta_{\rm sum}}]$ 
and for some $\gamma \in [0,\lambda]$, we have 
\begin{align}
& \tilde{\Omega}^{(\lambda)}(p)
= \tilde{\zeta}(\lambda)=\tilde{\zeta}(0)+ \tilde{\zeta}^{\prime}(0)\lambda
  +\frac{\tilde{\zeta}^{\prime\prime}(0)}{2}\irr{\lambda^2}
  +\frac{\tilde{\zeta}^{\prime\prime\prime}(\irr{\gamma})}{6}\irr{\lambda^3}
\notag\\
&=\lambda {\rm E}_p\left[\omega_{p}(\underline{X})\right]
   +\frac{\irr{\lambda^2}}{2}{\rm Var}_{p}
   \left[\omega_{p}(\underline{X})\right]
   +\frac{\irr{\lambda^3}}{6}
    \biggl(\frac{ {\rm d}^3\tilde{\Omega}^{(\beta)}(p)}
          {{\rm d}\beta^3}\biggr)_{\beta=\gamma}
\notag\\
&\leq \lambda \tilde{\Psi}_{\max} 
   +\frac{\irr{\lambda^2}}{2}{\rm Var}_{p}
   \left[\omega_{p}(\underline{X})\right] 
  +\frac{\irr{\lambda^3}}{6}
     \biggl(\frac{ {\rm d}^3\tilde{\Omega}^{(\lambda)}(p)}
          {{\rm d}\lambda^3}\biggr)_{\beta=\gamma}.
\label{eqn:AsZzz} 
\end{align}
For $\lambda \in [0,\frac{1}{2\eta_{\rm sum}}]$, set 
\begin{align*}
&c^{(\lambda)}\defeq 
\max_{\scs p\in \tilde{P}:
\atop{\scs 
      \tilde{\Omega}^{(\lambda)}(p)=
      \tilde{\Omega}_{\max}^{(\lambda)}(p)
      }
}
\max_{ \gamma \in [0,\lambda]}
\left|\biggl(
\frac{1}{6}
    \frac{ {\rm d}^3\tilde{\Omega}^{(\beta)}(p)}
          {{\rm d}\beta^3}\biggr)_{\beta=\gamma}
\right|.
\end{align*}
Furthermore, set
\begin{align*}
c^{(+)}=&c^{(+)}(\eta_{\rm sum})
\defeq \max_{\lambda \in [0,\frac{1}{2\eta_{\rm sum}}]}
c^{(\lambda)}.
\end{align*}
Note that such $c^{(+)}(\eta_{\rm sum})$ exists since 
$|\tilde{\zeta}^{\prime\prime\prime}(\irr{\gamma})|$ is bounded 
for $\gamma \in [0,\frac{1}{2\eta_{\rm sum}}].$ 
For each $\lambda \in [0,\frac{1}{2\eta_{\rm sum}}]$, we let 
$p^{(\lambda)}_{\rm opt}$ denote the probability distribution
that attains the maximum of 
${\rm Var}_p[\omega_p(\underline{X})]$ 
subject to 
$\tilde{\Omega}^{(\lambda)}(p)$
$=\tilde{\Omega}_{\max}^{(\lambda)}$.
By definition we have 
\begin{align}
& \tilde{\Omega}^{(\lambda)}(p^{(\lambda)}_{\rm opt})=
\tilde{\Omega}_{\max}^{(\lambda)},
\label{eqn:DistA}
\\
& {\rm Var}_{p^{(\lambda)}_{\rm opt}}
[\omega_{ p^{(\lambda)}_{\rm opt} }(\underline{X})]=\rho^{(\lambda)}. 
\label{eqn:DistB}
\end{align}
For each $\lambda \in [0,\frac{1}{2\eta_{\rm sum}}]$,
we choose $p=p^{(\lambda)}_{\rm opt}$ 
in (\ref{eqn:AsZzz}). Then for any 
$\lambda \in [0,\frac{1}{2\eta_{\rm sum}}]$, we have the 
following chain of inequalities:
\begin{align*}
& \tilde{\Omega}_{\max}^{(\lambda)}
\MEq{a}\tilde{\Omega}^{(\lambda)}(p^{(\lambda)}_{\rm opt})
\notag\\
&\MLeq{b} \lambda \tilde{\Psi}_{\max} 
   +\frac{\irr{\lambda^2}}{2} \rho^{(\lambda)}
   +\frac{\irr{\lambda^3}}{6}
    \biggl(\frac{ {\rm d}^3\tilde{\Omega}^{(\beta)}
    (p^{(\lambda)}_{\rm opt})}
          {{\rm d}\lambda^3}\biggr)_{\beta=\gamma}
\notag\\
&\MLeq{c} \lambda \tilde{\Psi}_{\max} 
   +\frac{\irr{\lambda^2}}{2} \rho^{(\lambda)}
   +\irr{\lambda^3}c^{(\lambda)}
\notag\\
& \MLeq{d} \lambda \tilde{\Psi}_{\max} 
   +\frac{\irr{\lambda^2}}{2} \rho^{(+)}
   +\irr{\lambda^3}c^{(+)}.
\end{align*}
Step (a) follows from (\ref{eqn:DistA}).
Step (b) follows from (\ref{eqn:AsZzz}) and (\ref{eqn:DistB}).
Step (c) follows from the definition of $c^{(\lambda)}$.
Step (d) follows from the definitions of 
$\rho^{(+)}$ and $c^{(+)}$.
\hfill\IEEEQED

{\it Proof of Property \ref{pr:pro1b} part c):}
Let $\tau$ be a small negative number.
Fix any $p \in \tilde{\cal P}$.  
By the Taylor expansion of 
$\tilde{\Omega}^{(\tau)}(p)$
with respect to $\tau$ around $\tau=0$,
we have that
for any $(p, \tau)\in$
$\tilde{\cal P}\times$ $[-\frac{1}{2\mu_{\rm sum}},0]$ 
and for some $\gamma \in [\tau,0]$, we have 
\begin{align}
& \tilde{\Omega}^{(\tau)}(p)
= \tilde{\zeta}(\tau)=\tilde{\zeta}(0)+ \tilde{\zeta}^{\prime}(0)\tau
  +\frac{\tilde{\zeta}^{\prime\prime}(0)}{2}\irr{\tau^2}
  +\frac{\tilde{\zeta}^{\prime\prime\prime}(\irr{\gamma})}{6}\irr{\tau^3}
\notag\\
&=\tau {\rm E}_p\left[\omega_{p}(\underline{X})\right]
   +\frac{\irr{\tau^2}}{2}{\rm Var}_{p}
   \left[\omega_{p}(\underline{X})\right]
   +\frac{\irr{\tau^3}}{6}
    \biggl(\frac{ {\rm d}^3\tilde{\Omega}^{(\beta)}(p)}
          {{\rm d}\beta^3}\biggr)_{\beta=\gamma}
\notag\\
&\leq \tau \tilde{\Psi}_{\min}
   +\frac{\irr{\tau^2}}{2}{\rm Var}_{p}
   \left[\omega_{p}(\underline{X})\right]
   +\frac{\irr{\tau^3}}{6}
    \biggl(\frac{ {\rm d}^3\tilde{\Omega}^{(\beta)}(p)}
          {{\rm d}\beta^3}\biggr)_{\beta=\gamma}.
\label{eqn:AsZzzP}
\end{align}
Putting $\tau=-\lambda$ for $\lambda>0$ in (\ref{eqn:AsZzzP}), we have  
that for any $\lambda \in [0,\frac{1}{2\mu_{\rm sum}}]$,  
\begin{align}
 \tilde{\Omega}^{(-\lambda)}(p)
\leq & -\lambda \tilde{\Psi}_{\min}
    +\frac{ \irr{\lambda^2}}{2} {\rm Var}_{p}
    \left[\omega_{p}(\underline{X})\right]
\notag\\
    & -\frac{\irr{\lambda^3}}{6}
    \biggl(\frac{ {\rm d}^3\tilde{\Omega}^{(\beta)}(p)}
          {{\rm d}\beta^3}\biggr)_{\beta=\gamma}.
\label{eqn:AsZzzPxx} 
\end{align}
For $\lambda \in [0,\frac{1}{2\mu_{\rm sum}}]$, set 
\begin{align*}
&c^{(-\lambda)}\defeq 
\max_{\scs p \in \tilde{P}:
\atop{\scs 
      \tilde{\Omega}^{(-\lambda)}(p)=
      \tilde{\Omega}_{\max}^{(-\lambda)}(p)
      }
}
\max_{ \gamma \in [-\lambda,0]}
\left|\biggl(
\frac{1}{6}
    \frac{ {\rm d}^3\tilde{\Omega}^{(\beta)}(p)}
          {{\rm d}\beta^3}\biggr)_{\beta=\gamma}
\right|.
\end{align*}
Furthermore, set
\begin{align*}
c^{(-)}=&c^{(-)}(\mu_{\rm sum})
\defeq \max_{\lambda \in [0,\frac{1}{2\mu_{\rm sum}}]}
c^{(-\lambda)}.
\end{align*}
Note that such $c^{(-)}(\mu_{\rm sum})$ exists since 
$|\tilde{\zeta}^{\prime\prime\prime}(\irr{\gamma})|$ is bounded 
for $\gamma \in [-\frac{1}{2\mu_{\rm sum}},0].$ 
For each $\lambda \in [0,\frac{1}{2\mu_{\rm sum}}]$, we let 
$p^{(-\lambda)}_{\rm opt}$ denote the probability distribution
that attains the maximum of 
${\rm Var}_p[\omega_p(\underline{X})]$ 
subject to 
$\tilde{\Omega}^{(-\lambda)}(p)$
$=\tilde{\Omega}_{\max}^{(-\lambda)}$.
By definition we have 
\begin{align}
& \tilde{\Omega}^{(-\lambda)}(p^{(-\lambda)}_{\rm opt})=
\tilde{\Omega}_{\max}^{(-\lambda)},
\label{eqn:DistAz}
\\
& {\rm Var}_{p^{(-\lambda)}_{\rm opt}}
[\omega_{ p^{(-\lambda)}_{\rm opt} }(\underline{X})]=\rho^{(-\lambda)}. 
\label{eqn:DistBz}
\end{align}
For each $\lambda \in [0,\frac{1}{2\mu_{\rm sum}}]$,
we choose $p=p^{(-\lambda)}_{\rm opt}$ in (\ref{eqn:AsZzzPxx}). 
Then for any 
$\lambda \in [0,\frac{1}{2\mu_{\rm sum}}]$, we have the 
following chain of inequalities:
\begin{align*}
& \tilde{\Omega}_{\max}^{(-\lambda)}
\MEq{a}\tilde{\Omega}^{(-\lambda)}(p^{(-\lambda)}_{\rm opt})
\notag\\
&\MLeq{b} -\lambda \tilde{\Psi}_{\min} 
   +\frac{\irr{\lambda^2}}{2} \rho^{(-\lambda)}
   -\frac{\irr{\lambda^3}}{6}
    \biggl(\frac{ {\rm d}^3\tilde{\Omega}^{(\beta)}(p^{(-\lambda)}_{\rm opt})}
          {{\rm d}\beta^3}\biggr)_{\beta=\gamma}
\notag\\
&\MLeq{c} -\lambda \tilde{\Psi}_{\min} 
    +\frac{\irr{\lambda^2}}{2} \rho^{(-\lambda)}
    +\irr{\lambda^3}c^{(-\lambda)}
\notag\\
&
\MLeq{d}-\lambda \tilde{\Psi}_{\min} 
   +\frac{\irr{\lambda^2}}{2} \rho^{(-)}
   +\irr{\lambda^3}c^{(-)}.
\end{align*}
Step (a) follows from (\ref{eqn:DistAz}).
Step (b) follows from (\ref{eqn:AsZzzPxx}) and (\ref{eqn:DistBz}).
Step (c) follows from the definition of $c^{(-\lambda)}$.
Step (d) follows from the definitions of 
$\rho^{(-)}$ and $c^{(-)}$.
\hfill\IEEEQED

}

\newcommand{\ApdaAACa}{
\subsection{
Proofs of Properties \ref{pr:pro1} and \ref{pr:pro1b} Part d)
} 
\label{sub:ApdaAACa}


In this appendix we derive the bound (\ref{eqn:AiSs0}) 
in Property \ref{pr:pro1} part d) and 
the bound (\ref{eqn:AiSs0z}) in Property \ref{pr:pro1b} 
part d). We first prepare two lemmas necessary for 
the proof. For $l=1,2,\cdots, L_3$, we set
\begin{align*}
& E_l\defeq
{\rm E}_q\left[\frac{ p^{\frac{\theta\kappa_{\rm sum}}{1-\theta\alpha}}
_{X_{{A}_{2l}} | X_{{A}_{2l-1}}}
         (x_{{A}_{2l}} | x_{{A}_{2l-1}}) }
     { q^{\frac{\theta\kappa_{\rm sum}}{1-\theta\alpha}}
_{X_{{A}_{2l}} | X_{{A}_{2l-1}}}
         (x_{{A}_{2l}} | x_{{A}_{2l-1}}) }\right].
\end{align*}
For $l=1,2,\cdots, L_4$, we set
\begin{align*}
& F_l\defeq
{\rm E}_q\left[\frac{ p^{-\frac{\theta\nu_{\rm sum}}{1+\theta\alpha}}
_{X_{{B}_{2l}} | X_{{B}_{2l-1}}}
         (x_{{B}_{2l}} | x_{{B}_{2l-1}}) }
     { q^{-\frac{\theta\nu_{\rm sum}}{1+\theta\alpha}}
_{X_{{B}_{2l}} | X_{{B}_{2l-1}}}
         (x_{{B}_{2l}} | x_{{B}_{2l-1}}) }\right].
\end{align*}
Then we have the following lemma.
\begin{lm}\label{lm:ErxP}
When 
$$
0< \theta \leq \frac{1}{\alpha+\kappa_{\rm sum}}
\mbox{ or equivalent to }
0< \frac{\theta\kappa_{\rm sum}}{1-\theta \alpha}\leq 1,
$$
we have $E_l \leq 1$ for $l=1,2,\cdots, L_3$.
When 
$$
-\frac{1}{\alpha+\nu_{\rm sum}}\leq \theta <0
\mbox{ or equivalent to }
0< \frac{-\theta\nu_{\rm sum}}{1+\theta \alpha}\leq 1,
$$
we have $F_l \leq 1$ for $l=1,2,\cdots, L_4$.
\end{lm}


{\it Proof: } 
When 
$$
0< \frac{\theta\kappa_{\rm sum}}{1-\theta \alpha}\leq 1,
$$
we apply H\"older's inequality to $E_l$ to obtain 
\begin{align*}
E_l&
\leq \left(
{\rm E}_q 
\left[\frac{p_{X_{{A}_{2l}} | X_{{A}_{2l-1}}}
        (x_{{A}_{2l}} | x_{{A}_{2l-1}})}
     {q_{X_{{A}_{2l}} | X_{{A}_{2l-1}}}
         (x_{{A}_{2l}} | x_{{A}_{2l-1}})}
\right]
\right)^{\frac{\theta\kappa_{\rm sum}}{1-\theta\alpha}}
=1,
\end{align*}
When
$$
0< \frac{-\theta\nu_{\rm sum}}{1+\theta \alpha}\leq 1,
$$
we apply H\"older's inequality to $F_l$ to obtain 
\begin{align*}
F_l&
\leq \left(
{\rm E}_q\left[\frac{ p_{X_{{B}_{2l}} | X_{{B}_{2l-1}}}
         ({X}_{{B}_{2l}} | {X}_{{B}_{2l-1}}) }
     { q_{X_{{B}_{2l}} | X_{{B}_{2l-1}}}
         ({X}_{{B}_{2l}} | {X}_{{B}_{2l-1}}) }
\right]
\right)^{-\frac{\theta\nu_{\rm sum}}{1+\theta\alpha}}
=1,
\end{align*}
completing the proof.
\hfill \IEEEQED 

We set 
\begin{align*}
&\exp\left\{ \hat{\Omega}^{(\alpha)}(p,q)\right\}
\defeq
{\rm E}_p\HUgebl 
\exp\left\{\frac{1}{\alpha} \omega_p(\underline{X}) \right\}
\nonumber\\
&\qquad \times 
\prod_{l=1}^{L_4}
      \left( 
\frac{ q_{X_{{B}_{2l}} | X_{{B}_{2l-1}}}
         (x_{{B}_{2l}} | x_{{B}_{2l-1}}) }
     { p_{X_{{B}_{2l}} | X_{{B}_{2l-1}}}
         (x_{{B}_{2l}} | x_{{B}_{2l-1}}) }
     \right)^{\frac{\nu_l}{\alpha} }
\HUgebr,
\nonumber\\
&\exp\left\{ \check{\Omega}^{(\alpha)}(p,q)\right\}
\defeq
{\rm E}_p\HUgebl 
\exp\left\{\frac{-1}{\alpha} \omega_p(\underline{X}) \right\}
\nonumber\\
&\qquad \times 
\prod_{l=1}^{L_3}
      \left( 
\frac{ q_{X_{{A}_{2l}} | X_{{A}_{2l-1}}}
         (x_{{A}_{2l}} | x_{{A}_{2l-1}}) }
     { p_{X_{{A}_{2l}} | X_{{A}_{2l-1}}}
         (x_{{A}_{2l}} | x_{{A}_{2l-1}}) }
     \right)^{\frac{\kappa_l}{\alpha} }
\HUgebr.
\end{align*}
Note that 
$\exp\left\{  \hat{\Omega}^{(\alpha)}(p,q)\right\}$
and $\exp\left\{ \check{\Omega}^{(\alpha)}(p,q)\right\}$
can also be written as
\begin{align}
&\exp\left\{ \hat{\Omega}^{(\alpha)}(p,q)\right\}
=
{\rm E}_q\HUgebl 
\frac{p(\underline{X})} {q(\underline{X})}
\exp\left\{\frac{1}{\alpha} \omega_p(\underline{X}) \right\}
\nonumber\\
&\qquad \times 
\prod_{l=1}^{L_4}
      \left( 
\frac{ q_{X_{{B}_{2l}} | X_{{B}_{2l-1}}}
         (x_{{B}_{2l}} | x_{{B}_{2l-1}}) }
     { p_{X_{{B}_{2l}} | X_{{B}_{2l-1}}}
         (x_{{B}_{2l}} | x_{{B}_{2l-1}}) }
     \right)^{\frac{\nu_l}{\alpha} }
\HUgebr, 
\label{eqn:ddAsal}\\
& \exp \left\{ \check{\Omega}^{(\alpha)}(p,q) \right\}
=
{\rm E}_q \HUgebl 
 \frac{ p(\underline{X})} {q(\underline{X})}
\exp\left\{ \frac{-1}{\alpha} \omega_p(\underline{X}) \right\}
\nonumber\\
&\qquad \times 
\prod_{l=1}^{L_3}
      \left( 
\frac{ q_{X_{{A}_{2l}} | X_{{A}_{2l-1}}}
         (x_{{A}_{2l}} | x_{{A}_{2l-1}}) }
     { p_{X_{{A}_{2l}} | X_{{A}_{2l-1}}}
         (x_{{A}_{2l}} | x_{{A}_{2l-1}}) }
     \right)^{\frac{\kappa_l}{\alpha} }
\HUgebr. 
\label{eqn:ddAsalz}
\end{align}
Then we have the following lemma.
\begin{lm}\label{lm:ErxPzz}
Fix any 
$q \in \tilde{\cal P}$ and $p=p^{(q)}=\varphi(p)$.
For any $\alpha > \nu_{\rm sum}$, we have   
$$
\hat{\Omega}^{(\alpha)}(p,q)
\leq \left(1-\frac{\nu_{\rm sum}}{\alpha}\right)
\tilde{\Omega}^{(\frac{1}{\alpha-\nu_{\rm sum}})}(p).
$$
For any $\alpha > \kappa_{\rm sum}$, we have   
$$
\check{\Omega}^{(\alpha)}(p,q)
\leq \left(1-\frac{\kappa_{\rm sum}}{\alpha}\right)
\tilde{\Omega}^{(\frac{-1}{\alpha-\kappa_{\rm sum}})}(p).
$$
\end{lm}

{\it Proof: }
We consider the case where
$$
1-\sum_{l=1}^{L_4}\frac{\nu_l}{\alpha}=1-\frac{\nu_{\rm sum}}{\alpha}>0. 
$$
The above condition is equivalent to $\alpha > \nu_{\rm sum}$. 
Then we have the following: 
\begin{align*}
&\exp\left\{ \hat{\Omega}^{(\alpha)}(p,q)\right\}
={\rm E}_p\HUgebl 
\left(\exp\left\{
\frac{1}{\alpha-{\nu_{\rm sum}}}\omega_p(\underline{X}) \right\}
\right)^{1-\frac{\nu_{\rm sum}}{\alpha}} 
\nonumber\\
&\qquad \times 
\prod_{l=1}^{L_4}
      \left( 
\frac{ q_{X_{{B}_{2l}} | X_{{B}_{2l-1}}}
         (x_{{B}_{2l}} | x_{{B}_{2l-1}}) }
     { p_{X_{{B}_{2l}} | X_{{B}_{2l-1}}}
         (x_{{B}_{2l}} | x_{{B}_{2l-1}}) }
     \right)^{\frac{\nu_l}{\alpha} }
\HUgebr
\nonumber\\
& \MLeq{a}
\left( 
{\rm E}_p\left[
\exp\left\{\frac{1}{\alpha-\nu_{\rm sum}}\omega_p(\underline{X})\right\}
\right]\right)^{1-\frac{\nu_{\rm sum}}{\alpha}}
\nonumber\\
&\qquad \times 
\prod_{l=1}^{L_4}
      \left(
{\rm E}_p\left[ 
\frac{ q_{X_{{B}_{2l}} | X_{{B}_{2l-1}}}
         (x_{{B}_{2l}} | x_{{B}_{2l-1}}) }
     { p_{X_{{B}_{2l}} | X_{{B}_{2l-1}}}
         (x_{{B}_{2l}} | x_{{B}_{2l-1}}) }
\right]
     \right)^{\frac{\kappa_l}{\alpha}}
\nonumber\\
&=\exp\left\{\left(1- \frac{ \nu_{\rm sum} } {\alpha} \right)
     \tilde{\Omega}^{(\frac{1}{\alpha-\nu_{\rm sum}})}(p)\right\}.
\end{align*}
Step (a) follows from H\"older's inequality.
We next consider the case where
$$
1-\sum_{l=1}^{L_3}\frac{\kappa_l}{\alpha}=1-\frac{\kappa_{\rm sum}}{\alpha}>0. 
$$
The above condition is equivalent to $\alpha > \kappa_{\rm sum}$. 
Then we have the following: 
\begin{align*}
&\exp\left\{ \check{\Omega}^{(\alpha)}(p,q)\right\}
={\rm E}_p\HUgebl 
\left(\exp\left\{ 
\frac{-1}{\alpha-\kappa_{\rm sum}}\omega_p(\underline{X}) \right\}
\right)^{1-\frac{\kappa_{\rm sum}}{\alpha}} 
\nonumber\\
&\qquad \times 
\prod_{l=1}^{L_3}
      \left( 
\frac{ q_{X_{{A}_{2l}} | X_{{A}_{2l-1}}}
         (x_{{A}_{2l}} | x_{{A}_{2l-1}}) }
     { p_{X_{{A}_{2l}} | X_{{A}_{2l-1}}}
         (x_{{A}_{2l}} | x_{{A}_{2l-1}}) }
     \right)^{\frac{\kappa_l}{\alpha} }
\HUgebr
\nonumber\\
& \MLeq{a}
\left( 
{\rm E}_p\left[
\exp\left\{\frac{-1}{\alpha-\kappa_{\rm sum}}
\omega_p(\underline{X})\right\}
\right]\right)^{1-\frac{\kappa_{\rm sum}}{\alpha}}
\nonumber\\
&\qquad \times 
\prod_{l=1}^{L_3}
      \left(
{\rm E}_p\left[ 
\frac{ q_{X_{{A}_{2l}} | X_{{A}_{2l-1}}}
         (x_{{A}_{2l}} | x_{{A}_{2l-1}}) }
     { p_{X_{{A}_{2l}} | X_{{A}_{2l-1}}}
         (x_{{A}_{2l}} | x_{{A}_{2l-1}}) }
\right]
     \right)^{\frac{\kappa_l}{\alpha}}
\nonumber\\
&=\exp\left\{\left(1- \frac{ \kappa_{\rm sum} } {\alpha} \right)
     \tilde{\Omega}^{(\frac{-1}{\alpha-\kappa_{\rm sum}})}(p)\right\}.
\end{align*}
Step (a) follows from H\"older's inequality.
\hfill \IEEEQED 



{\it Proof of the bound (\ref{eqn:AiSs0}) 
in Property \ref{pr:pro1} Part d):}
On $\exp\{$ ${\Omega}^{(\theta,\alpha)}(q)\}$ we have 
the following: 
\begin{align}
&\exp \left\{ {\Omega}^{(\theta,\alpha)}(q)\right\}
={\rm E}_q 
\left[
\exp\left\{\theta \omega_q(\underline{X}) \right\}
\left\{
\frac{p(\underline{X})}{q(\underline{X})}
\right\}^{\theta\alpha}
\right]
\nonumber\\
&\MEq{a}{\rm E}_q 
\left[
\left\{
\frac{p(\underline{X})}{q(\underline{X})}
\right\}^{\theta\alpha}
\exp\left\{\theta \omega_p(\underline{X}) \right\}
\right.
\nonumber\\
&\quad \times 
\prod_{l=1}^{L_4}
      \left( 
\frac{ q_{X_{{B}_{2l}} | X_{{B}_{2l-1}}}
         (x_{{B}_{2l}} | x_{{B}_{2l-1}}) }
     { p_{X_{{B}_{2l}} | X_{{B}_{2l-1}}}
         (x_{{B}_{2l}} | x_{{B}_{2l-1}}) }
     \right)^{\theta \nu_l}
\nonumber\\
&\quad \times
\left.
\prod_{l=1}^{L_3}
      \left(
\frac{ p_{X_{{A}_{2l}} | X_{{A}_{2l-1}}}
         (x_{{A}_{2l}} | x_{{A}_{2l-1}}) }
     { q_{X_{{A}_{2l}} | X_{{A}_{2l-1}}}
         (x_{{A}_{2l}} | x_{{A}_{2l-1}}) }
      \right)^{\theta \kappa_l}
\right].
\label{eqn:SddEaaa}
\end{align}
Step (a) follows from Assumption \ref{asm:AssmTwo}. 
We choose $\theta>0$ so that 
$$0< \theta \leq \frac{1}{\alpha+\kappa_{\rm sum}}.$$
For this choice of $\theta$ and (\ref{eqn:SddEaaa}), 
we have the following chain of inequalities: 
\begin{align}
&\exp \left\{ {\Omega}^{(\theta,\alpha)}(q)\right\}
={\rm E}_q 
\HUgebl
\Hugel
\frac{p(\underline{X})}{q(\underline{X})}
\exp\left\{\frac{1}{\alpha} \omega_p(\underline{X}) \right\}
\nonumber\\
&\quad \times 
\prod_{l=1}^{L_4}
      \left( 
\frac{ q_{X_{{B}_{2l}} | X_{{B}_{2l-1}}}
         (x_{{B}_{2l}} | x_{{B}_{2l-1}}) }
     { p_{X_{{B}_{2l}} | X_{{B}_{2l-1}}}
         (x_{{B}_{2l}} | x_{{B}_{2l-1}}) }
     \right)^{\frac{\nu_l}{\alpha} }
\Huger^{\theta\alpha}
\nonumber\\
&\quad \times
\left.
\prod_{l=1}^{L_3}
      \left(
\frac{ p^{\frac{\theta\kappa_{\rm sum}}{1-\theta\alpha}}
_{X_{{A}_{2l}} | X_{{A}_{2l-1}}}
         (x_{{A}_{2l}} | x_{{A}_{2l-1}}) }
     { q^{\frac{\theta\kappa_{\rm sum}}{1-\theta\alpha}}
_{X_{{A}_{2l}} | X_{{A}_{2l-1}}}
         (x_{{A}_{2l}} | x_{{A}_{2l-1}}) }
      \right)^{\frac{\kappa_l}{\kappa_{\rm sum}}(1-\theta\alpha)}
\right]
\nonumber\\
&\MLeq{a}\HUgecl
{\rm E}_q\HUgebl \frac{p(\underline{X})}{q(\underline{X})}
\exp\left\{\frac{1}{\alpha} \omega_p(\underline{X}) \right\}
\nonumber\\
&\quad \times 
\prod_{l=1}^{L_4}
      \left( 
\frac{ q_{X_{{B}_{2l}} | X_{{B}_{2l-1}}}
         (x_{{B}_{2l}} | x_{{B}_{2l-1}}) }
     { p_{X_{{B}_{2l}} | X_{{B}_{2l-1}}}
         (x_{{B}_{2l}} | x_{{B}_{2l-1}}) }
     \right)^{\frac{\nu_l}{\alpha} }
\HUgebr
\HUgecr^{\theta\alpha}
\nonumber\\
&\quad \times 
\prod_{l=1}^{L_3}
\left({\rm E}_q\left[\frac{ p^{\frac{\theta\kappa_{\rm sum}}{1-\theta\alpha}}
_{X_{{A}_{2l}} | X_{{A}_{2l-1}}}
         (x_{{A}_{2l}} | x_{{A}_{2l-1}}) }
     { q^{\frac{\theta\kappa_{\rm sum}}{1-\theta\alpha}}
_{X_{{A}_{2l}} | X_{{A}_{2l-1}}}
         (x_{{A}_{2l}} | x_{{A}_{2l-1}}) }\right]
\right)^{\frac{\kappa_l}{\kappa_{\rm sum}}(1-\theta\alpha)} 
\nonumber\\
\newcommand{\Zass}{
&
\left(
{\rm E}_q 
\left[
\frac{p_{Y|X}^{\bar{\beta}\frac{\alpha}{\bar{\alpha}}}(Y|X)}
     {q^{\bar{\beta}\frac{\alpha}{\bar{\alpha}}}_{Y|X}(Y|X)}
\right]
\right)^{\bar{\alpha}}
\nonumber\\
}
&\MEq{b}\exp\left\{\theta \alpha \hat{\Omega}^{(\alpha)}(p,q)\right\}
\prod_{l=1}^{L_3}
[E_l]^{\frac{\kappa_l}{\kappa_{\rm sum}}(1-\theta\alpha)}
\nonumber\\ 
& \MLeq{c}\exp\left\{\theta \alpha \hat{\Omega}^{(\alpha)}(p,q)\right\}
\nonumber\\
&\MLeq{d}
\exp\left\{\theta(\alpha-\nu_{\rm sum})
\tilde{\Omega}^{(\frac{1}{\alpha-\nu_{\rm sum}})}(p)\right\}.
\nonumber
\end{align}
Step (a) follows from H\"older's inequality.
Step (b) follows from (\ref{eqn:ddAsal}).
Step (c) follows from Lemma \ref{lm:ErxP}.
Step (d) follows from Lemma \ref{lm:ErxPzz}.
\hfill \IEEEQED 

{\it Proof of the bound (\ref{eqn:AiSs0z}) 
in Property \ref{pr:pro1b} Part d):}
On $\exp\{$ ${\Omega}^{(\theta,-\alpha)}(q)\}$ we 
have the following: 
\begin{align}
&\exp \left\{ {\Omega}^{(\theta,-\alpha)}(q)\right\}
={\rm E}_q 
\left[
\exp\left\{\theta \omega_q(\underline{X}) \right\}
\left\{
\frac{p(\underline{X})}{q(\underline{X})}
\right\}^{-\theta\alpha}
\right]
\nonumber\\
&\MEq{a}{\rm E}_q 
\left[
\left\{
\frac{p(\underline{X})}{q(\underline{X})}
\right\}^{-\theta\alpha}
\exp\left\{\theta \omega_p(\underline{X}) \right\}
\right.
\nonumber\\
&\quad \times 
\prod_{l=1}^{L_3}
      \left( 
\frac{ q_{X_{{A}_{2l}} | X_{{A}_{2l-1}}}
         (x_{{A}_{2l}} | x_{{A}_{2l-1}}) }
     { p_{X_{{A}_{2l}} | X_{{A}_{2l-1}}}
         (x_{{A}_{2l}} | x_{{A}_{2l-1}}) }
     \right)^{\theta\kappa_l}
\nonumber\\
&\quad \times
\left.
\prod_{l=1}^{L_4}
      \left(
\frac{ p_{X_{{B}_{2l}} | X_{{B}_{2l-1}}}
         (x_{{B}_{2l}} | x_{{B}_{2l-1}}) }
     { q_{X_{{B}_{2l}} | X_{{B}_{2l-1}}}
         (x_{{B}_{2l}} | x_{{B}_{2l-1}}) }
      \right)^{\theta \nu_l}
\right].
\label{eqn:SddEr}
\end{align}
Step (a) follows from Assumption \ref{asm:AssmTwo}. 
We choose $\theta<0$ so that 
$$ \frac{-1}{\alpha+\nu_{\rm sum}} \leq \theta < 0.$$
For this choice of $\theta$ and (\ref{eqn:SddEr}), 
we have the following chain of inequalities: 
\begin{align}
&\exp \left\{ {\Omega}^{(\theta,-\alpha)}(q)\right\}
={\rm E}_q 
\HUgebl
\Hugel
\frac{p(\underline{X})}{q(\underline{X})}
\exp\left\{\frac{-1}{\alpha} \omega_p(\underline{X}) \right\}
\nonumber\\
&\quad \times 
\prod_{l=1}^{L_3}
      \left( 
\frac{ q_{X_{{A}_{2l}} | X_{{A}_{2l-1}}}
         (x_{{A}_{2l}} | x_{{A}_{2l-1}}) }
     { p_{X_{{A}_{2l}} | X_{{A}_{2l-1}}}
         (x_{{A}_{2l}} | x_{{A}_{2l-1}}) }
     \right)^{\frac{\kappa_l}{\alpha} }
\Huger^{-\theta\alpha}
\nonumber\\
&\quad \times
\left.
\prod_{l=1}^{L_4}
      \left(
\frac{ p^{-\frac{\theta\nu_{\rm sum}}{1+\theta\alpha}}
_{X_{{B}_{2l}} | X_{{B}_{2l-1}}}
         (x_{{B}_{2l}} | x_{{B}_{2l-1}}) }
     { q^{-\frac{\theta\nu_{\rm sum}}{1+\theta\alpha}}
_{X_{{B}_{2l}} | X_{{B}_{2l-1}}}
         (x_{{B}_{2l}} | x_{{B}_{2l-1}}) }
      \right)^{\frac{\nu_l}{\nu_{\rm sum}}(1+\theta\alpha)}
\right]
\nonumber\\
&\MLeq{a}\HUgecl
{\rm E}_q\HUgebl \frac{p(\underline{X})}{q(\underline{X})}
\exp\left\{\frac{-1}{\alpha} \omega_p(\underline{X}) \right\}
\nonumber\\
&\quad \times 
\prod_{l=1}^{L_3}
      \left( 
\frac{ q_{X_{{A}_{2l}} | X_{{A}_{2l-1}}}
         (x_{{A}_{2l}} | x_{{A}_{2l-1}}) }
     { p_{X_{{A}_{2l}} | X_{{A}_{2l-1}}}
         (x_{{A}_{2l}} | x_{{A}_{2l-1}}) }
     \right)^{\frac{\kappa_l}{\alpha} }
\HUgebr
\HUgecr^{-\theta\alpha}
\nonumber\\
&\quad \times 
\prod_{l=1}^{L_4}
\left({\rm E}_q\left[
 \frac{ p^{-\frac{\theta\nu_{\rm sum}}{1+\theta\alpha}}
_{X_{{B}_{2l}} | X_{{B}_{2l-1}}}
         (x_{{B}_{2l}} | x_{{B}_{2l-1}}) }
      { q^{-\frac{\theta\nu_{\rm sum}}{1+\theta\alpha}}
_{X_{{B}_{2l}} | X_{{B}_{2l-1}}}
         (x_{{B}_{2l}} | x_{{B}_{2l-1}}) }\right]
\right)^{\frac{\nu_l}{\nu_{\rm sum}}(1+\theta\alpha)} 
\nonumber\\
\newcommand{\Zass}{
&
\left(
{\rm E}_q 
\left[
\frac{p_{Y|X}^{\bar{\beta}\frac{\alpha}{\bar{\alpha}}}(Y|X)}
     {q^{\bar{\beta}\frac{\alpha}{\bar{\alpha}}}_{Y|X}(Y|X)}
\right]
\right)^{\bar{\alpha}}
\nonumber\\
}
&\MEq{b}\exp\left\{-\theta\alpha\check{\Omega}^{(\alpha)}(p,q)\right\}
\prod_{l=1}^{L_4}
[F_l]^{\frac{\nu_l}{\nu_{\rm sum}}(1+\theta\alpha)}
\nonumber\\ 
& \MLeq{c}\exp\left\{-\theta\alpha\check{\Omega}^{(\alpha)}(p,q)\right\}
\nonumber\\
&\MLeq{d}
\exp\left\{-\theta(\alpha-\kappa_{\rm sum})
\tilde{\Omega}^{(\frac{-1}{\alpha-\kappa_{\rm sum}})}(p)\right\}.
\nonumber
\end{align}
Step (a) follows from H\"older's inequality.
Step (b) follows from (\ref{eqn:ddAsalz}).
Step (c) follows from Lemma \ref{lm:ErxP}.
Step (d) follows from Lemma \ref{lm:ErxPzz}.
\hfill \IEEEQED 

\newcommand{\ZapSSS}{
where we set
\begin{align*}
& E_l\defeq
{\rm E}_q\left[\frac{ p^{\frac{\theta\kappa_{\rm sum}}{1-\theta\alpha}}
_{X_{{A}_{2l}} | X_{{A}_{2l-1}}}
         (x_{{A}_{2l}} | x_{{A}_{2l-1}}) }
     { q^{\frac{\theta\kappa_{\rm sum}}{1-\theta\alpha}}
_{X_{{A}_{2l}} | X_{{A}_{2l-1}}}
         (x_{{A}_{2l}} | x_{{A}_{2l-1}}) }\right].
\end{align*}
Step (b) follows from H\"older's inequality. From (\ref{eqn:EddxSSS}), 
we can see that it suffices to show $E_l \leq 1,$ 
$l=1,2,\cdots,L_3$ to complete the proof. When 
$$
\frac{\theta\kappa_{\rm sum}}{1-\theta \alpha}\leq 1
\mbox{ or equivalent to }
0< \theta \leq \frac{1}{\alpha+\kappa_{\rm sum}},
$$
we apply H\"older's inequality to $E_l$ to obtain 
\begin{align*}
E_l&=
{\rm E}_q\left[\frac{ p^{\frac{\theta\kappa_{\rm sum}}{1-\theta\alpha}}
_{X_{{A}_{2l}} | X_{{A}_{2l-1}}}
         (x_{{A}_{2l}} | x_{{A}_{2l-1}}) }
     { q^{\frac{\theta\kappa_{\rm sum}}{1-\theta\alpha}}
_{X_{{A}_{2l}} | X_{{A}_{2l-1}}}
         (x_{{A}_{2l}} | x_{{A}_{2l-1}}) }\right]
\notag\\
&
\leq \left(
{\rm E}_q 
\left[\frac{p_{X_{{A}_{2l}} | X_{{A}_{2l-1}}}
        (x_{{A}_{2l}} | x_{{A}_{2l-1}})}
     {q_{X_{{A}_{2l}} | X_{{A}_{2l-1}}}
         (x_{{A}_{2l}} | x_{{A}_{2l-1}})}
\right]
\right)^{\frac{\theta\kappa_{\rm sum}}{1-\theta\alpha}}
=1,
\end{align*}
completing the proof.
\hfill \IEEEQED 
}

}

\section*{\empty}
\appendix

\ApdaProOne
\ApdaAAC
\ApdaAACa

\end{document}